\documentclass[12pt,preprint]{aastex}

\newif\ifrev
\revfalse
\newif\ifdel
\delfalse
\usepackage{xcolor}
\usepackage{ulem}
\usepackage{censor}

\newlength\nextcharwidth
\makeatletter
\renewcommand\@cenword[1]{%
  \setlength{\nextcharwidth}{\widthof{#1}}%
  \censorrule{\nextcharwidth}%
  \kern -\nextcharwidth%
  #1}
\makeatother
\newcommand{\delref}[1]{\ifdel\textcolor{blue}{\censorruledepth=.55ex\censor{#1}}\else\fi}
\newcommand{\rev}[1]{\ifrev\textcolor{magenta}{#1}\else{#1}\fi}
\newcommand{\del}[1]{\ifdel\textcolor{blue}{\sout{#1}}\else\fi}

\newcommand{\Ms}{$M_\odot$}
\newcommand{\Msyr}{$M_\odot$ yr$^{-1}$}
\newcommand{\Msyrkpcsq}{$M_\odot$ yr$^{-1}$ kpc$^{-2}$}

\newcommand{\Ha}{H${\alpha}$}

\newcommand{\akari}{\textit{AKARI}}
\newcommand{\galex}{\textit{GALEX}}
\newcommand{\galexfull}{\textit{Galaxy Evolution Explorer}}
\newcommand{\herschel}{\textit{Herschel}}

\newcommand{\Rv}{$R_V$}

\newcommand{\sfrsd}{$\Sigma_{\mathrm{SFR,UV}}$}
\newcommand{\zdD}{$z_d^{\mathrm{thick}}/D_{25}$}
\newcommand{\sfruv}{SFR$_{\mathrm{UV}}$}
\newcommand{\Mex}{$M_{ext}$}
\newcommand{\Mst}{$M_*$}

\usepackage[english]{babel}
\usepackage[utf8]{inputenc}
\usepackage{amsmath}
\usepackage{graphicx}
\usepackage[colorinlistoftodos]{todonotes}

\title{Ultraviolet Radiative Transfer Modeling of Nearby Galaxies with Extraplanar Dusts}

\author{Jong-Ho Shinn\altaffilmark{1} and Kwang-Il Seon\altaffilmark{1,2,3}}
\email{jhshinn@kasi.re.kr}
\altaffiltext{1}{Korea Astronomy and Space Science Institute, 776 Daeduk-daero, Yuseong-gu, Daejeon, 305-348, the Republic of Korea}
\altaffiltext{2}{Astronomy and Space Science Major, Korea University of Science and Technology, 217 Gajeong-ro, Yuseong-gu, Daejeon, 305-350, the Republic of Korea}
\altaffiltext{3}{Department of Astrophysical Sciences, Princeton University, Princeton NJ 08544, USA}

\date{\today}

\begin{document}

\begin{abstract}
In order to examine their relation to the host galaxy, the extraplanar dust of six nearby galaxies are modeled, employing a three dimensional Monte Carlo radiative transfer code. 
The targets are from the highly-inclined galaxies that show dust-scattered ultraviolet halos, and the archival \galexfull{} FUV band images were fitted with the model.
The \del{extraplanar dust is}\rev{observed images are in general} well reproduced by two dust layers and one light-source layer, whose vertical and radial distributions have exponential profiles.
We obtained several important physical parameters, such as star formation rate (\sfruv), face-on optical depth, and scale-heights.
\del{The extraplanar dust mass was found to have high correlations with both \sfruv{} and galactic stellar mass (\Mst) which scales with the galactic luminosity.
These high correlations indicate the scaling relation among the three quantities, which constraints the pushing force to scale with \Mst{} in order to balance with the gravity.
The galactic radiation pressure, one of the pushing force candidates, satisfies this scaling property, since it scales with the luminosity.
We considered other dynamical mechanisms, however they seem to be relatively minor due to their localized activities which is in contrast to the galactic radiation's characteristic---global and plane-parallel propagation away from the galactic plane.}
\rev{
Three galaxies (NGC 891, NGC 3628, and UGC 11794) show clear evidence for the existence of extraplanar dust layer.
However, it is found that the rest three targets (IC 5249, NGC 24, and NGC 4173) do not necessarily need a thick dust disk to model the ultraviolet (UV) halo, because its contribution is too small and the UV halo may be caused by the wing part of the \galex{} point spread function.
This indicates that the galaxy samples reported to have UV halos may be contaminated by galaxies with negligible extraplanar (halo) dust.
The galaxies showing evidence of the extraplanar dust layer fall within a narrow range on the scatter plots between physical parameters such as \sfruv{} and extraplanar dust mass.
Several mechanisms possible to produce the extraplanar dust are discussed.
}
We also found a hint that the extraplanar dust scale-height might not be much different from the polycyclic aromatic hydrocarbon \textit{emission} characteristic height.
\end{abstract}


\section{Introduction \label{intro}}
Galaxies evolve circulating their mediums between disks and halos, through stellar winds and radiation, supernova explosions, dynamic interactions between galaxies, gravitational attraction towards the galactic plane, star formation by gravitational collapse, etc \citep[cf.][]{Finkbeiner(2014)Science_346_905}.
These circulating mediums thus embed important information regarding how the energy is transported and how the raw material for star formation are distributed within the galaxies and their surroundings.
Dust is one of such mediums, and it plays several important roles.
For instance, it attenuates the stellar light through the scattering and absorption processes, and re-radiates the absorbed light in longer wavelengths to transform the spectral energy distribution of the galaxy \citep[cf.][]{Conroy(2013)ARA&A_51_393}.

The dust that is located above the galactic plane (say, $z\gtrsim$ a few percent of the galactic diameter), i.e.~the extraplanar dust, is especially important in several aspects.
First, its existence means that the dust is somehow elevated against, or accreted by, the galactic plane's gravity \citep[cf.][]{Howk(2012)inproc}.
Second, it may reveal the evolutionary connection between the star-forming galaxies and the starburst galaxies, where the star-formation activity can supply the force defying the galactic plane's gravity.
Third, it may hold some hints on the dust supplying channel to the circumgalactic medium and the intergalactic medium, where the existence of dust was reported \citep[e.g.][]{Menard(2010)MNRAS_405_1025,Zaritsky(1994)AJ_108_1619,McGee(2010)MNRAS_405_2069,Chelouche(2007)ApJ_671_L97}.
Fourth, its existence is intriguing in consideration of the possibility that the extraplanar warm ionized medium (or diffuse ionized gas, \citealt{Haffner(2009)RvMP_81_969}) could be substantially contaminated by the dust-scattered \Ha{} emission \cite[cf.][]{Seon(2012)ApJ_758_109}.

The extraplanar dust has been observed for many edge-on spiral galaxies.
The edge-on galaxies reveal the extraplanar dust through the extinction of background stellar light \citep{Howk(1997)AJ_114_2463,Howk(1999)AJ_117_2077,Howk(2000)AJ_119_644,Alton(2000)A&AS_145_83,Rossa(2003)A&A_406_505,Thompson(2004)AJ_128_662}, the mid-infrared continuum \citep{Burgdorf(2007)ApJ_668_918}, and ultraviolet \rev{(UV)} scattered light \citep{Hodges-Kluck(2014)ApJ_789_131,Seon(2014)ApJ_785_L18}.
Indirect evidences from the emission of polycyclic aromatic hydrocarbon (PAH) were also reported \citep{McCormick(2013)ApJ_774_126,Irwin(2006)A&A_445_123}.
However, most of previous studies analyzed the data without a physical modeling for the galaxy, and therefore it was hard to extract direct links between physical parameters.
For example, the correlation study using the extraplanar PAH emission characteristic height, presented by \cite{McCormick(2013)ApJ_774_126}, could be misled by the dilution of radiation field away from the galactic plane, because the PAH emission is \textit{excited} by the absorption of starlight \citep{Draine(2003)ARA&A_41_241,Allamandola(1989)ApJS_71_733,Leger(1984)A&A_137_L5} and hence the emission does not directly correspond to the actual PAH amount.
It is also true that a scale-height of the scattered light above the galactic plane is not the same as the actual scale-height of the extraplanar dust layer.

Recently, from the radiative transfer simulation of the \del{ultraviolet}\rev{UV} images, \cite{Seon(2014)ApJ_785_L18} showed that the extraplanar dust of NGC 891 can properly be modeled by adding a geometrically-thick dust layer.
They derived several important physical parameters such as star formation rate (SFR) and dust scale-height, and listed two primary elevation mechanisms for the extraplanar dust: stellar radiation pressure or (magneto)hydrodynamic flows \citep{Howk(1997)AJ_114_2463,Ferrara(1990)A&A_240_259,Greenberg(1987)Nature_327_214}.
We here extend their approach to the highly-inclined spiral galaxies with dust-scattered \del{ultraviolet}\rev{UV} halos, discovered by \cite{Hodges-Kluck(2014)ApJ_789_131}, in order to investigate if there is any relation between the properties of extraplanar dust and host galaxy.
\rev{
Six targets were analyzed, and three of them found to need no additional thick dust disk in modeling the UV halos.
The rest three that need the thick dust disk fall within a narrow range over the scatter plots between several physical parameters such as \sfruv{} and extraplanar dust mass.
}
\del{Six targets were analyzed and the extraplanar dust mass was found to have a high correlation with both SFR and galactic stellar mass (i.e.~galactic luminosity).}

\section{Target, Instrument, and Data Reduction \label{obs-red}}
We are interested in modeling and characterizing the extraplanar dust with several physical parameters.
\cite{Seon(2014)ApJ_785_L18} showed that the extraplanar dust of NGC 891 can be described by the exponential distributions of starlight source and dust.
Among the spiral galaxies studied by \cite{Hodges-Kluck(2014)ApJ_789_131}, we selected the galaxies that seem to be properly modeled with the exponential distributions.
Those galaxies with ring structures or prominent spiral arms were excluded.
Table \ref{tbl-tg} lists the six selected targets.
Their distance, major axis size, stellar mass, and SFR range $\sim5-75$ Mpc, $\sim10-50$ kpc, $\sim(0.1-7.0)\times10^{10}$ \Ms, and $\sim0.1-5.0$ \Msyr, respectively.

\begin{figure}
\center{
\includegraphics[scale=1.0]{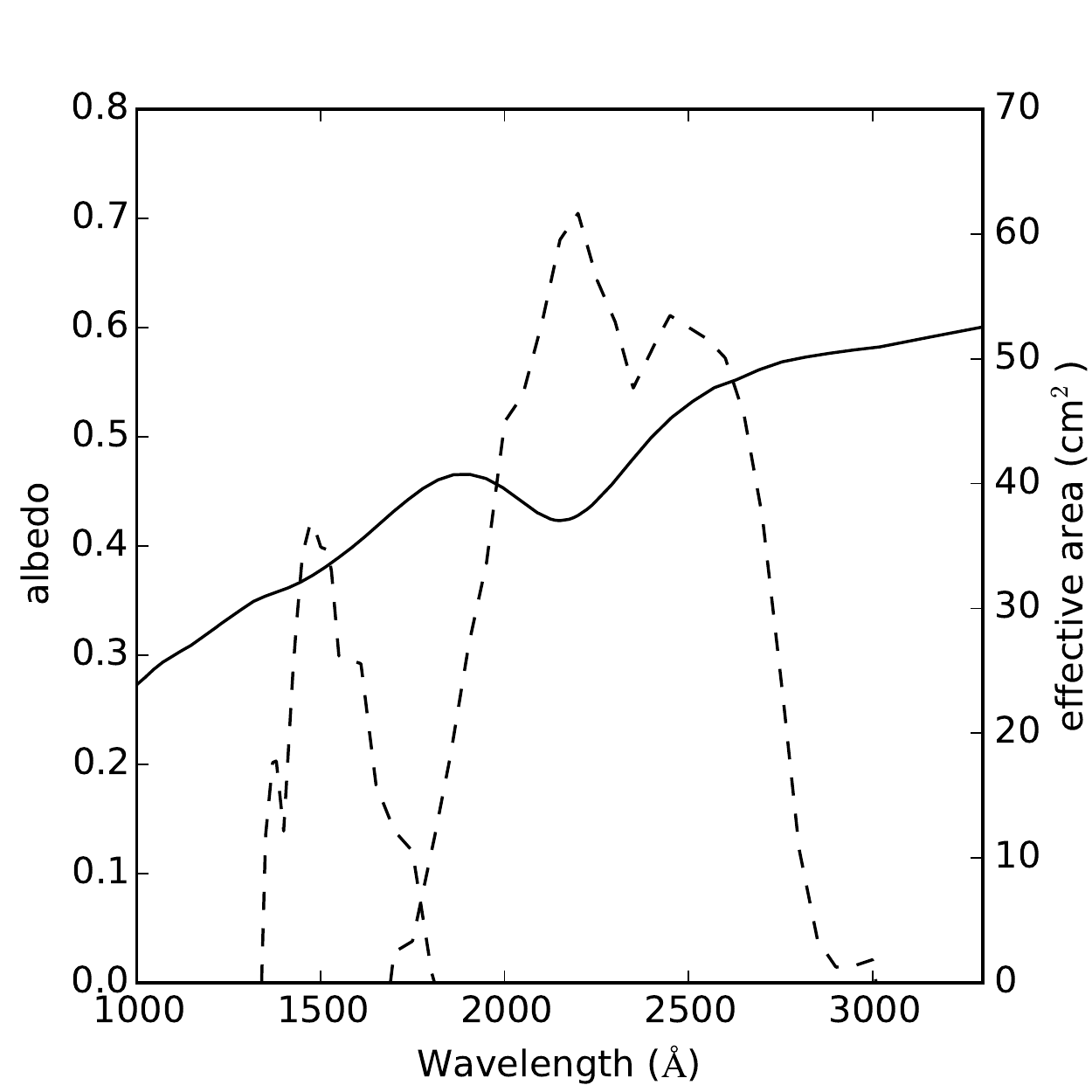}
}
\caption{The albedo curve from the \Rv=3.1 Milky Way dust model of \cite{Draine(2003)ARA&A_41_241} (solid) and the \galex{} FUV/NUV effective area curves (dashed). \label{fig-band}}
\end{figure}

We employed the archival data\footnote{GR6/GR7; \url{http://galex.stsci.edu/GR6/}} of \galexfull\ (\galex; \citealt{Martin(2005)ApJ_619_L1}).
\galex\ was a satellite mission that performed all-sky imaging and spectroscopic survey in two \del{ultraviolet}\rev{UV} bands: FUV ($\sim$1350--1750 \AA) and NUV ($\sim$1750--2750 \AA).
Table \ref{tbl-galex} shows more detailed information on \galex.
The archive provides fully reduced data, and the data reduction pipeline is described in \cite{Morrissey(2007)ApJS_173_682}.
We used the intensity maps, for which no background subtraction was applied during the data reduction.
The FUV images were solely used to model the extraplanar dust, because the FUV band has a narrower band-width than the NUV band, and does not include the 2175 \AA{} bump which might cause the sudden variation of dust grain albedo (Figure \ref{fig-band}, cf.~\citealt{Draine(2003)ARA&A_41_241}).
Furthermore, the FUV image has a smaller number of foreground point sources than the NUV image, which makes the point-source masking process easier.
Table \ref{tbl-exp} shows the exposure times of individual targets at each band; the NUV exposure time is also shown for reference.

In order to render the intensity maps suitable for two dimensional fitting, we applied additional processes as follows.
First, we masked point-like sources, employing the SExtractor \citep{Bertin(1996)A&AS_117_393} output provided by the archive.
Other remaining point-like sources were masked manually, and then the blank sky and the target galaxy were left intact.
Second, we rotated the masked image for the major axis of the galaxy to be placed horizontal, and cropped the rotated image.
These rotating and cropping were implemented using IRAF \citep{Tody(1986)inproc,Tody(1993)inproc}.
Finally, we subtracted the linear sky background that have a gradient along the galaxy's minor axis direction.
This background was derived by fitting the averaged vertical profile of the cropped image with a linear function.
During the fitting, we excluded the central part where the features from the galaxy and extraplanar diffuse sources are not negligible.
Figure \ref{fig-data} shows the resultant FUV images and their vertical profiles, which reveal the existence of \del{ultraviolet}\rev{UV} halos.

\section{Analysis and Results \label{ana-res}}
\subsection{Fitting the \galex{} FUV images with a three dimensional radiative transfer model}
We are interested in the physical properties of the extraplanar dust layer, such as the scale-height and the optical depth.
Several physical parameters of the extraplanar dust layer are derived by exploiting three dimensional radiative transfer model to fit the \galex{} FUV image processed as described in section \ref{obs-red}.
We used a three dimensional Monte Carlo radiative transfer code of \cite{Seon(2014)ApJ_785_L18}, which models multiple scattering of photons in a cylindrical coordinate system.
It uses the ``forced first scattering'' technique to enhance the calculation efficiency in optically thin media and the ``peel-off'' technique to generate an image towards the observer (see \citealt{Steinacker(2013)ARA&A_51_63} for more about the computational acceleration techniques). 
The code also employs a fast voxel traversal algorithm \citep{Amanatides(1987)inproc}, as implemented in \cite{Seon(2015)JKAS_48_57}.

The model configuration was the same as in \cite{Seon(2014)ApJ_785_L18}.
We adopted the axisymmetric distribution of dust and starlight source, and their radial and vertical profiles follow the exponential curves\footnote{For the vertical profile, it may be hard to distinguish the exponential type from the power-law type, unless the signal-to-noise ratio is high enough at the distant area from the galactic plane. We note that a power-law \textit{radial} profile was adopted in \cite{Menard(2010)MNRAS_405_1025} and \cite{Hodges-Kluck(2014)ApJ_789_131}.}.
The dust component has two layers which share the same physical parameters except the scale-height and optical depth: one has a smaller scale-height (geometrically thin), while the other has a larger one (geometrically thick).
In brief, the following equations represent the dust and starlight distributions.
\begin{eqnarray}
\kappa(r,z)&=& 
\left\{ 
\begin{array}{ll} \label{eq-kappa}
\kappa_0^{\mathrm{thin}}\,\mathrm{exp}\left(-\frac{r}{h_d}-\frac{|z|}{z_d^{\mathrm{thin}}} \right)+\kappa_0^{\mathrm{thick}}\,\mathrm{exp}\left(-\frac{r}{h_d}-\frac{|z|}{z_d^{\mathrm{thick}}} \right) ,\,\mathrm{for}\, r\le R_d,\, z\le Z_{d} \\
0,\,\mathrm{for}\, r>R_d,\, z>Z_d 
\end{array} 
\right. \\
I(r,z)&=&
\left\{
\begin{array}{ll}
I_0\,\mathrm{exp}\left(-\frac{r}{h_s}-\frac{|z|}{z_s} \right),\,\mathrm{for}\, r\le R_s \\
0,\,\mathrm{for}\, r>R_s 
\end{array}
\right.
\end{eqnarray}
$\kappa(r,z)$ and $I(r,z)$ are the distributions of extinction coefficient and starlight source, respectively.
We set the starlight wavelength to be the effective wavelength of \galex{} FUV band, 1538.6 \AA{} \citep{Morrissey(2007)ApJS_173_682}.
$r$ and $z$ are the cylindrical coordinates.
$z_d^{\mathrm{thin}}$, $z_d^{\mathrm{thick}}$, and $z_s$ are the scale-heights of geometrically thin and thick dust layers, and starlight sources, respectively.
$h_d$ and $h_s$ are the scale-lengths of the dust layer and starlight, respectively.
$\kappa_0^{\mathrm{thin}}$ and $\kappa_0^{\mathrm{thick}}$ are the extinction coefficients at the center of the galaxy ($r=0$, $z=0$) of the thin and thick dust layers, respectively.
Note that the optical depth along the symmetric axis is $\tau=\tau^{\mathrm{thin}}+\tau^{\mathrm{thick}}=2\kappa_0^{\mathrm{thin}}z_d^{\mathrm{thin}}+2\kappa_0^{\mathrm{thick}}z_d^{\mathrm{thick}}$.
$R_d$ and $R_s$ are the truncation radii of the dust layers and the starlight, respectively.
$Z_d$ is the truncation height of the dust distribution.

For the light scattering calculation, we used the Henyey-Greenstein phase function \citep{Henyey(1941)ApJ_93_70} and adopted the anisotropy parameter and albedo from the \Rv=3.1 Milky Way dust model of \cite{Draine(2003)ARA&A_41_241} and \cite{Weingartner(2001)ApJ_548_296}.
The modeled image was convolved with the instrumental point spread function (PSF), before comparing with the processed \galex{} FUV image.
In the present paper, the observed images were not folded about the minor axis during the comparison, whereas those were folded in \cite{Seon(2014)ApJ_785_L18}.

The model fitting was performed through $\chi^2$ minimization.
We adopted an equal error for all image pixels to give the equal weights for the disk and halo during the fitting, and the error was estimated from the pixel value fluctuation at backgrounds.
Since $z_d^{\mathrm{thick}}$ is our main interest, we surveyed several fixed parameter space of ($z_d^{\mathrm{thin}}$, $z_d^{\mathrm{thick}}$; for $z_d^{\mathrm{thin}} < z_d^{\mathrm{thick}}$) and searched for the best fit parameters.
$z_d^{\mathrm{thin}}$ ranges from 0.025 kpc to 0.275$-$0.375 kpc with a 0.05 kpc step, while $z_d^{\mathrm{thick}}$ ranges from 0.1 kpc to 2.6$-$3.6 kpc with a 0.5 kpc step.
Table \ref{tbl-fit} shows the best-fit model parameters for individual targets, and Figure \ref{fig-fit} shows the corresponding model images.
The $\chi^2$ differences between the minimum and its surrounding ($z_d^{\mathrm{thin}}$, $z_d^{\mathrm{thick}}$) are much larger than one.
Thus, the 1-$\sigma$ uncertainties of scale-heights are much smaller than their step sizes.

\begin{figure}
\center{
\includegraphics[scale=1.0,angle=0]{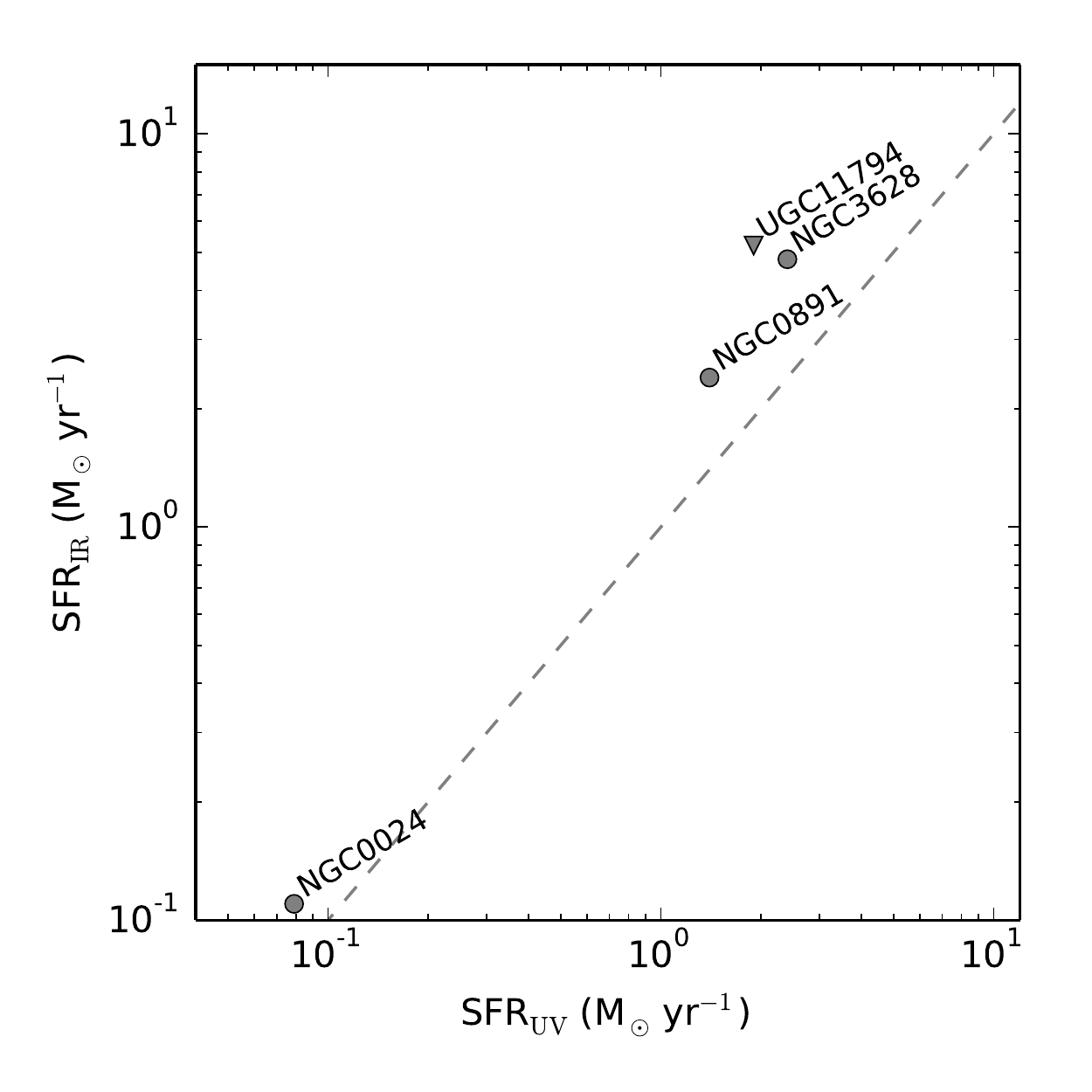}
}
\caption{Comparison of Star Formation Rate. The $\mathrm{SFR_{UV}}$ is from the \galex{} image fitting (Table \ref{tbl-fit}), while the $\mathrm{SFR_{IR}}$ is from the IR luminosity (Table \ref{tbl-tg}). The dashed line indicates the 1-to-1 ratio. This plot shows only the targets that have the $\mathrm{SFR_{IR}}$ information (cf.~Table \ref{tbl-tg}). \rev{The downward triangle indicates the upper limit of $\mathrm{SFR_{IR}}$.} \label{fig-sfr}}
\end{figure}

The best-fit models have $z_d^{\mathrm{thin}}$ of \del{$\sim0.1-0.3$}\rev{$\sim0.1-0.4$} kpc and $z_d^{\mathrm{thick}}$ of \del{$\sim0.6-2.1$}\rev{$\sim0.6-3.1$} kpc.
This $z_d^{\mathrm{thin}}$ range is comparable to the values from previous optical and near infrared (NIR) studies on the galactic disks with a single dust layer \citep{Xilouris(1998)A&A_331_894,Xilouris(1999)A&A_344_868,Alton(2004)A&A_425_109}.
$\tau_B^{\mathrm{thin}}$ and $\tau_B^{\mathrm{thick}}$ range \del{$\sim0.2-1.2$}\rev{$\sim0.4-1.2$} and \del{$\sim0.1-0.6$}\rev{$\sim0.01-1.0$}, respectively.
$\tau_B^{\mathrm{thin}}$ is greater than $\tau_B^{\mathrm{thick}}$ for all targets except \del{IC 5249}\rev{NGC 24}.
The $\tau_B^{\mathrm{thin}}$ range is comparable to the values from previous optical and NIR studies \citep{Xilouris(1998)A&A_331_894,Xilouris(1999)A&A_344_868,Alton(2004)A&A_425_109}.
$z_s$ ranges $\sim0.1-0.2$ kpc, which is lower than the values from previous optical studies \citep[$\sim0.2-0.5$ kpc,][]{Xilouris(1998)A&A_331_894,Xilouris(1999)A&A_344_868}.
This may reflect the fact that early-type stars, which are bright in \del{ultraviolet}\rev{UV}, have a lower scale-height than late-type stars \citep[e.g.][]{Bahcall(1980)ApJS_44_73,Wainscoat(1992)ApJS_83_111}.
SFR was estimated from the intrinsic \del{ultraviolet}\rev{UV} luminosity, employing the Starburst99 \citep{Leitherer(1999)ApJS_123_3}.
The Salpeter initial mass function was assumed in calculating the UV luminosity per unit SFR.
We also assumed that our sample galaxies are in a steady and continuous star-forming phase.
The derived SFRs (\sfruv) are comparable \del{and consistent} with the SFRs derived from the infrared luminosity (SFR$_{\mathrm{IR}}$, Figure \ref{fig-sfr})\footnote{\rev{SFR$_{\mathrm{IR}}$ is omitted for IC 5249 and NGC 4173, because there is no far-infrared data that can be used in the SFR estimation (see Table \ref{tbl-tg}). For example, they are not listed in the catalog of \herschel{} and \akari.}}\rev{, although \sfruv{} is a little lower than SFR$_{\mathrm{IR}}$}.
\rev{We note that the estimated \sfruv{} includes some uncertainties, in the sense that the Starburst99 model is based on several (probable) assumptions, for instance, in star formation history and metallicity, and the assumption we made in the radiative transfer modeling---smooth distribution of dust and light source---is a simplified representation of galaxies.}

We note that the obtained fitting parameters for NGC 891 are comparable in general to the $\tau_B^{\mathrm{thin}}=0.9$ case values of \cite{Seon(2014)ApJ_785_L18}.
This shows that our parameter search method over ($z_d^{\mathrm{thin}}$, $z_d^{\mathrm{thick}}$) grids returns a coherent output.
\del{IC 5249 is the only target whose $\tau_B^{\mathrm{thick}}$ is greater than $\tau_B^{\mathrm{thin}}$, and its}\rev{IC 5249's} $h_s$ is much larger than others\del{.
These peculiarities}\rev{, and this peculiarity} may be related with its very flat appearance (Fig.~\ref{fig-fit}a).
NGC 24 has the \del{lowest}\rev{highest} $\tau_B^{\mathrm{thick}}$\rev{, $z_d^{\mathrm{thick}}$ and lowest $h_d$, \sfruv,} and is the least edge-on.
\del{NGC 891 has the highest $z_d^{\mathrm{thick}}$ as does NGC 3628.}
NGC 3628 has the highest \del{$\tau_B^{\mathrm{thick}}$ and} \sfruv.
It has many local features, but the global shape is well reproduced by the model (Fig.~\ref{fig-fit}d).
NGC 4173 has the lowest $z_d^{\mathrm{thick}}$ \del{as does IC 5249, and has the lowest \sfruv}.
UGC 11794 has the smallest angular size, and hence the data quality is the lowest among our targets.
However, the fit results give reasonable parameters.

\subsubsection{Effects of radially extended PSF on the model fitting \label{ana-res-fit-psf}}
When observing astronomical objects, optical systems inevitably accompany scattered lights, which are characterized by the wing part of PSF.
It was recently claimed that a radially extended PSF should be used in the analyses, if the low surface-brightness features around bright sources are the targets of interest \citep{Sandin(2014)A&A_567_A97,Sandin(2015)A&A_577_A106}; \cite{Sandin(2014)A&A_567_A97} found that the PSF radius should be larger than 1.5 times the distance between the low surface-brightness feature and bright source.
Some of our targets have UV halos extended up to $\sim150''$ from the galactic plane (see Figure \ref{fig-data}), which is larger than the radius of the \galex{} standard PSF ($\sim90''$).
Therefore, we check if the extended wing part of PSF could have significant effects on the model fitting.

\begin{figure}
\center{
\includegraphics[scale=1.0,angle=0]{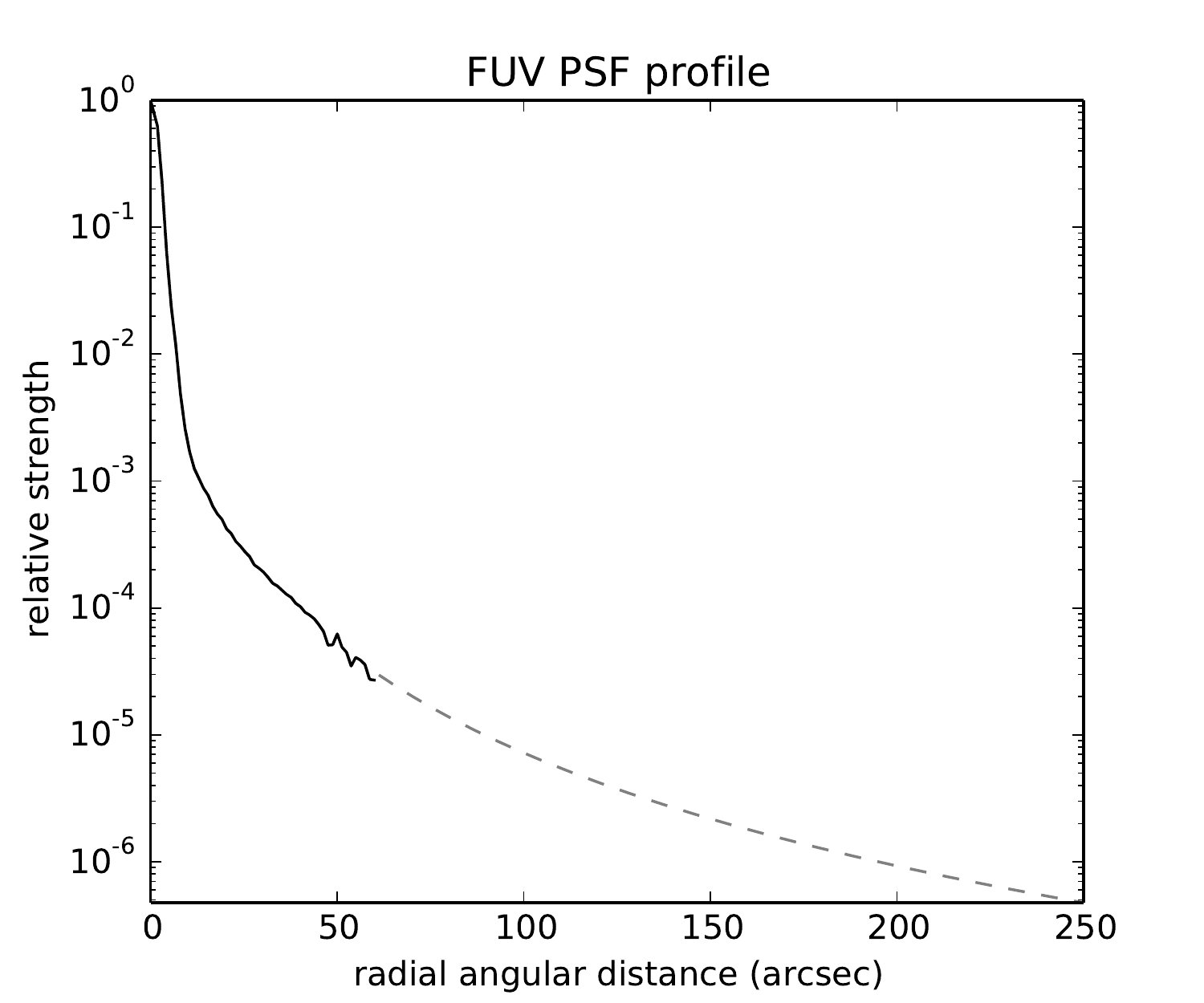}
}
\caption{The average radial profile of \galex{} FUV-band PSF (\textit{solid}) and the extrapolated wing part (\textit{dashed}). \label{fig-psf}}
\end{figure}

Since the extended PSF of \galex{} does not exist, we synthesized an extended PSF from the radial profile of standard PSF.
We fitted the average radial profile of standard PSF with two functions referring \cite{Racine(1996)PASP_108_699}: eq.~(6) of \cite[][the Moffat function]{Racine(1996)PASP_108_699} for the core and eq.~(7) of \cite{Racine(1996)PASP_108_699} for the wing.
The fitting was performed on the standard PSF cutting at the radius of $60''$ due to the larger fluctuation at higher radius.
Figure \ref{fig-psf} shows the fit result which extends up to the radius of $250''$.
The ratio of wing/core at the origin is about $2\times10^{-3}$, and the characteristic angular scales are $4.46''$ and $15.8''$ for the core and wing parts, respectively. 
From this extended radial profile, we made a synthetic, extended PSF whose radial extent reaches $250''$.

\rev{
In order to visualize the effects of extended PSF on the galactic halo, we made plots of profiles averaged along the horizontal axes, as seen in Figure \ref{fig-diff}.
The \textit{red-dashed} and \textit{magenta-solid} lines are the best-fit model profiles, convoluted with the \textit{extended} and \textit{standard} PSFs, respectively, excluding the thick dust disk component (see eq.~(\ref{eq-kappa})).
If the UV halo is caused by the extended PSF wing rather than the extraplanar dust, the \textit{red-dashed} line must significantly be higher than the \textit{magenta-solid} line at the region above the background noise upper bound (\textit{black-dotted} line in Figure \ref{fig-diff}).
We found that the \textit{red-dashed} lines are higher than the \textit{magenta-solid} line at the region \textit{below or marginally above} the noise upper bound for all targets.
This result seems to be due to that the most contribution of PSF wing part is already included in the standard PSF, whose radial extent is $\sim90''$, and that the background signal-to-noise ratio of \galex{} FUV data is not high enough to reveal the extended wing contribution.
}

\rev{
The vertical profiles of NGC 891, NGC 3628, and UGC 11794 clearly show the halo emission excess which cannot be explained with the PSF wing.
However, we found that the existence of thick dust disk is questionable for the rest three targets---IC 5249, NGC 24, and NGC 4173.
As Figure \ref{fig-diff} shows, for these targets, the \textit{magenta-solid} line is hardly distinguishable from the \textit{black-solid} line, the profile from the best-fit model including both thin and thick dust disks (see eq.~(\ref{eq-kappa})) and being convoluted with the standard PSF.
These targets have lower $\tau_B^{\mathrm{thick}}\times( h_d^{\mathrm{thick}})^2$ than other targets (see Table \ref{tbl-fit}), which indicates having lower amount of the extraplanar dust (see section \ref{ana-res-sfr}) and subsequently makes the two profiles indistinguishable.
In the case of NGC 24, its less edge-on appearance makes the difference even weaker.
These three targets were excluded from our further analyses.
}

\rev{
We here note that some of the galactic UV halos reported by \cite{Hodges-Kluck(2014)ApJ_789_131} may be explainable \textit{with the thin dust disk solely}.
The data profiles of IC 5249 and NGC 24 seen in Figure \ref{fig-diff} show hints of extended, decreasing wing-like features, but with intensities comparable to or lower than the background upper bound.
The feature for IC 5249 can be explained by the \textit{standard} PSF wing solely.
However, for NGC 24, the \textit{extended} PSF seems to reproduce the profile better.
It should be noted that the wing features caused by the PSF can be misidentified as extended UV halos, without the aid of detailed modeling that properly includes the PSF effects.
}

\del{
In order to visualize the effects of the extended PSF, we made plots of difference profiles averaged along the horizontal axis.
Figure \ref{fig-diff} shows two difference profiles: one is (best-fit model convoluted with the \textit{extended} PSF) $-$ (best-fit model convoluted with the \textit{standard} PSF), and the other is (data) $-$ (best-fit model convoluted with the \textit{standard} PSF).
The former falls within the noise scale of the latter.
This means that the difference that the extended PSF makes is indistinguishable from the data noise.
We thus think the effect of the extended PSF on the model fitting is negligible.
}

\subsection{Extraplanar dust and the host-galaxy radiation \label{ana-res-sfr}}
Since our main interest is to investigate the relation between \rev{properties of} the extraplanar dust \rev{and the host galaxy}\del{ properties and the host-galaxy radiation}, we first check whether or not there is any correlation between the extraplanar dust mass (\Mex) and \sfruv.
The extraplanar dust was defined to be the dust located outside a certain vertical distance ($z_x$) from the galactic plane.
We set $z_x$ to be 5\%{} of the major axis size ($D_{25}$); for example, NGC 891 has $z_x\simeq2$ kpc (see Table \ref{tbl-zxMext}).
\Mex{} is estimated as follows.
We integrate the extinction coefficient over the extraplanar volume, where $0 \le r \le R_d$ and $z_x \le z \le Z_d$ (cf.~eq.~(\ref{eq-kappa})).
This returns a value proportional to $\tau_B\,h_d^2$, which is converted to \Mex, using the extinction curve from the \Rv=3.1 Milky Way dust model of \cite{Draine(2003)ARA&A_41_241} and \cite{Weingartner(2001)ApJ_548_296}.

\begin{figure}
\center{
\includegraphics[scale=0.6,angle=0]{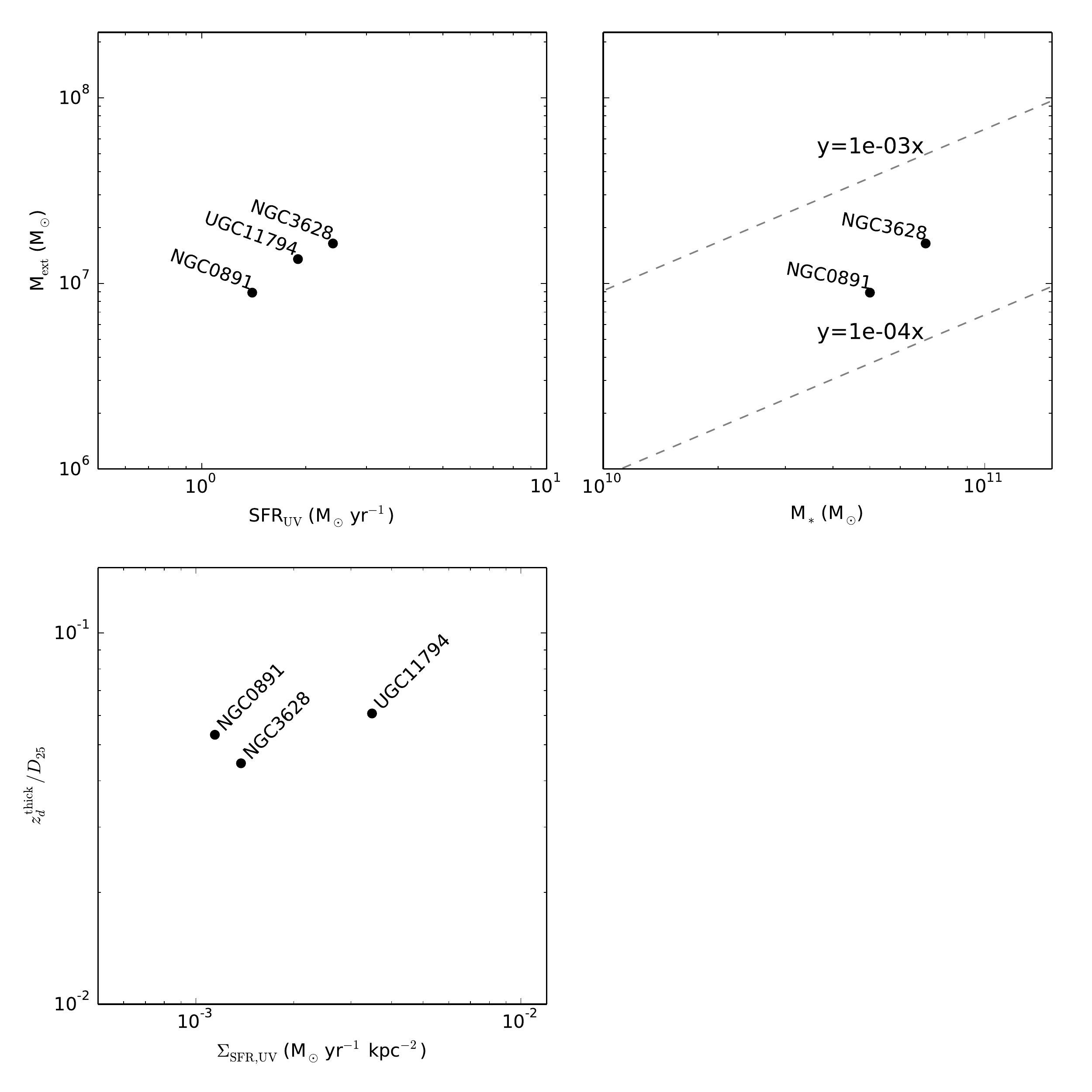}
}
\caption{Relation between Physical Parameters from the Model Fitting. \del{The numbers on the top side of each panel are the Pearson linear correlation coefficient ($r$) and the corresponding two-tailed $p$-value.} The scatter plots of the extraplanar dust mass (dust mass of $|z|>5\%\times D_{25}$) with the star formation rate (\textit{upper-left}) and the stellar mass (\textit{upper-right}). The stellar mass is from Table \ref{tbl-tg}. \del{The dashed-lines are a power-law fit to the data (\textit{upper-left}) and guide lines (\textit{upper-right}), respectively. The corresponding equations are shown in each panel.}\rev{The guide lines are shown in the \textit{upper-right} panel.} (\textit{lower-left}) The scatter plot between the star formation rate surface density and the ratio of the thick dust disk scale-height over the major axis size.
\label{fig-scatter}}
\end{figure}

The top-left panel of Figure \ref{fig-scatter} shows the scatter plot between \sfruv{} (Table \ref{tbl-fit}) and \Mex{} estimated as above (Table \ref{tbl-zxMext}).
\del{It shows a very high linear-correlation with a Pearson  correlation coefficient of 0.99 and a $p$-value of 0.0002.
The data show a power-law relation as below.}
\del{\left(\frac{\textrm{\Mex}}{\textrm{\Ms}}\right)=5.30\times10^6\,\left(\frac{\textrm{\sfruv}}{\textrm{\Msyr}}\right)^{1.34}}
For reference to galaxy luminosity, the scatter plot between \Mex{} and the galactic stellar mass (\Mst) is also plotted in the top-right panel of Figure \ref{fig-scatter}.
\Mex{} falls between $10^{-4}$\Mst{} and $10^{-3}$\Mst.
\del{It shows a correlation coefficient of 0.98 and a $p$-value of 0.0153.}
We here note that \Mex{} of NGC 891 ($8.91\times10^{6}$ \Ms{} for $|z|>1.97$ kpc, Table \ref{tbl-zxMext}) is $\sim40$ times larger than that estimated by \cite{Alton(2000)A&AS_145_83}, $2.1\times10^{5}$ \Ms{} for $|z|>0.68$ kpc.
This difference is because \cite{Alton(2000)A&AS_145_83} only modeled the geometrically thin dust disk.
Their dust scale-height of 0.31 kpc is similar to our $z_d^{\mathrm{thin}}=0.325$ kpc (Table \ref{tbl-fit}).

We check another relation between the obtained model parameters, which is thought to be scale-independent, referring to \cite{McCormick(2013)ApJ_774_126}.
The bottom-left panel of Figure \ref{fig-scatter} shows a scatter plot between the \sfruv{} surface density ($\Sigma_{\mathrm{SFR,UV}}$) and the ratio of $z_d^{\mathrm{thick}}/D_{25}$.
$\Sigma_{\mathrm{SFR,UV}}$ is calculated from SFR$_{\mathrm{UV}}$ (Table \ref{tbl-fit}) and $D_{25}$ (Table \ref{tbl-tg}).
\del{It shows a rather weak correlation with a Pearson correlation coefficient of 0.22 and a $p$-value of 0.6751.
We here note that our data have a narrow dynamic range of \sfrsd{} from $\sim7\times10^{-4}$ to $\sim4\times10^{-3}$ \Msyrkpcsq.}
The present results seem to fit into the relationship found by \cite{McCormick(2013)ApJ_774_126} over a much larger SFR surface density ($\Sigma_{\mathrm{SFR}}$) dynamic range, although they used the \textit{PAH-emission} characteristic height instead of the dust scale-height.

\rev{
We cannot draw any conclusive results about the relation between the aforesaid quantities, since the number of samples is too small (see section \ref{ana-res-fit-psf}).
However, note that those quantities fall within a narrow range for all three targets (see Figure \ref{fig-scatter}).
}

\section{Discussion \label{discu}}
\rev{
We found that three of our targets do not need any additional dust disk other than the typical thin dust disk, in modeling the galactic UV halo radiation reported by \cite{Hodges-Kluck(2014)ApJ_789_131} (see section \ref{ana-res-fit-psf}).
\cite{Hodges-Kluck(2014)ApJ_789_131} showed low and high correlations of the specific halo UV luminosity with the specific SFR and the host-galaxy UV luminosity, respectively. 
Our result suggests that the relations found by \cite{Hodges-Kluck(2014)ApJ_789_131} may include some contamination from the galaxies with negligible halo (extraplanar) dust.
This also demonstrates how the radiative transfer modeling can contribute to the interpretation of the galactic UV halo radiation.
}

\rev{
Since the sample number is too small, it would be premature to make a definite statement based on the scatter plots between different physical quantities (Figure \ref{fig-scatter}).
However, even with three targets, Figure \ref{fig-scatter} shows the quantities fall within a narrow range.
We also note that the three targets with no clear evidence of the thick dust layer have much lower \sfruv{} than the other targets (Table \ref{tbl-fit}).
These suggest that a common mechanism might regulate the behavior of extraplanar dust.
}

\del{
As seen from Figure \ref{fig-scatter}, \Mex{} and \sfruv{} show a strong linear correlation ($r=0.99,\,p=0.0002$), while \zdD{} and \sfrsd, which might be treated as scale-independent quantities, show a weak correlation ($r=0.22,\,p=0.6751$).
We cannot impart much meaning yet to the low correlation between \zdD{} and \sfrsd{}, because the dynamic range of \sfrsd{} is small ([\sfrsd]$_{max}$/[\sfrsd]$_{min}\sim6$) as noted in section \ref{ana-res-sfr}.
On the other hand, the dynamic range of \sfruv{} ([\sfruv]$_{max}$/[\sfruv]$_{min}$) is about 30, and the strong correlation between \Mex{} and \sfruv{} likely indicates a physical meaning.
\Mex{} also shows a high correlation ($r=0.98,\,p=0.0153$) with \Mst{} which scales with the galactic luminosity.
}

\del{
The strong linear correlation of \Mex{} with \sfruv{} and \Mst{} indicates that the three quantities share a scaling relation: the more \Mex{} is from the more massive galaxy, where SFR is also higher }\delref{\citep[cf.][]{Lehmer(2010)ApJ_724_559,Zahid(2013)ApJ_763_92}.}\del{
If the extraplanar dust is steadily located above the galactic plane, the scaling relation gives a constraint on the mechanism that elevates the dust away from the plane.
The steady location of extraplanar dust means the balance between the pushing force and the gravitational, pulling force of the plane; two forces could be in either equilibrium (static) or non-equilibrium (outflowing or accreting).
The pushing force should therefore scale with \Mst, since the plane's gravitational force scales with \Mst.
}

\del{The galactic radiation pressure satisfies this scaling property, because it is proportional to the galactic luminosity which scales with \Mst.}
\rev{First, we can point out the galactic radiation pressure as a plausible mechanism.}
This mechanism was previously studied by several authors \citep{Ferrara(1991)ApJ_381_137,Franco(1991)ApJ_366_443,Ferrara(1990)A&A_240_259,Barsella(1989)A&A_209_349,Greenberg(1987)Nature_327_214,Pecker(1974)A&A_35_7,Pecker(1972)A&A_18_253,Chiao(1972)MNRAS_159_361}.
They showed that dust grains can be elevated above the galactic plane and even escaped from the galaxy by the radiation pressure, considering two dust species (silicate and graphite) and diverse dust sizes under various galactic radiation fields and environments.
\rev{
There is a hint that the difference of galactic spectral shape might not significantly alter the strength of the radiation pressure, from the wavelength-averaged radiation pressure coefficient $Q_{pr}^*$ for Sb and Sc type galaxies \citep{Ferrara(1991)ApJ_381_137}.
If this is the case and our target galaxies share a similar grain size distribution, the main factor that determines \Mex{} would be the galactic luminosity itself, rather than the spectral shape of the radiation field.
}

\del{
It is not straightforward to pinpoint which one (\sfruv{} or the galactic luminosity) is the more significant factor for increasing \Mex{}, since \Mex{} has strong correlations with \textit{both} \sfruv{} and \Mst{} (i.e.~the galactic luminosity), and the galactic spectral shape varies among spiral galaxies }\delref{\citep[cf.][]{Dale(2009)ApJ_703_517,Rowan-Robinson(2005)AJ_129_1183,Yoshii(1988)ApJ_326_1}.}\del{ 
However, we found a hint that the difference of galactic spectral shape might not have a significant effect on the radiation pressure, from the wavelength-averaged radiation pressure coefficient $Q_{pr}^*$ for Sb and Sc type galaxies }\delref{\citep{Ferrara(1991)ApJ_381_137}.}\del{
If this is the case and our target galaxies share a similar grain size distribution, the main factor that determines \Mex{} would be the galactic luminosity itself, rather than the radiation strength at ultraviolet wavelengths (i.e.~\sfruv).
In this case, the strong correlation between \Mex{} and \sfruv{} is likely caused by the higher sensitivity of UV luminosity value (i.e.~\sfruv) to \Mst{} than to the galactic spectral shape.
We should be able to constrain the grain species and size distribution of the extraplanar dust, in order to clarify the \Mex-determining factor.
}

\cite{Howk(1997)AJ_114_2463} listed other elevating mechanisms such as hydrodynamical phenomena, magnetic field effects, dynamical instabilities.
They exemplified galactic fountain flows \citep{Bregman(1980)ApJ_236_577,Houck(1990)ApJ_352_506}, Parker instability by horizontal magnetic field \citep{Parker(1966)ApJ_145_811}, and vertical instability by non-axisymmetric potentials \citep{Binney(1981)MNRAS_196_455,Mulder(1984)A&A_134_158}.
\del{The galactic fountain originated from supernova explosions might scale with \Mst, since \Mst{} scales with SFR, and hence the number of massive stars.
The other two factors---magnetic field and non-axisymmetric potential---might scale with \Mst.}
However, we think that all three mechanisms have some difficulties in \rev{directly} producing the observed global-and-diffuse \del{ultraviolet}\rev{UV} halos \citep{Hodges-Kluck(2014)ApJ_789_131}, because their activities are likely to be local, not widespread over the disk.\footnote{Note that these three mechanisms might be applicable to the filamentary, vertical dust features discovered by \cite{Howk(1997)AJ_114_2463}.}
\rev{
In the case of galactic fountain, the hot gas which is shock-heated by supernova explosions may be able to spread dust over the galactic halo somewhat evenly due to their high mobility, if the dust dragging by the hot gas is strong enough.
}
The galactic radiation \del{has}\rev{may have some} advantage \del{over}\rev{in} producing the \rev{global-and-diffuse UV} halo, since the radiation propagates approximately plane-parallel away from the galactic plane. 
It should be noted that many UV bright features close to the galactic mid-plane were found to correlate with star-forming regions in the mid-plane, especially for NGC 891 \citep{Seon(2014)ApJ_785_L18}. 
These bright, scattered features might be due to the local mechanisms, as for the filamentary absorption features observed by \cite{Howk(1997)AJ_114_2463}.

\rev{
Another possible mechanism is the dust accretion from the circumgalactic or intergalactic medium.
The existence of dust in the circumgalactic and intergalactic medium has been known from previous studies \citep[e.g.][]{Menard(2010)MNRAS_405_1025,Zaritsky(1994)AJ_108_1619,McGee(2010)MNRAS_405_2069,Chelouche(2007)ApJ_671_L97}.
Therefore, if this external dust accretes to the host galaxy by the gravitational attraction  \cite[cf.][]{Oosterloo(2007)AJ_134_1019}, the extraplanar dust can be formed for the host galaxy.
However, it is not clear whether the accretion process can make the diffuse-and-global distribution of extraplanar dust, including the PAH molecules.
We here note that the diffuse UV halo radiation reported by \cite{Hodges-Kluck(2014)ApJ_789_131} can have some localized features unresolved by \galex{} and \textit{Swift}, whose spatial resolutions are $\sim2-5''$.
}

Other relations between the extraplanar dust and SFR can be seen from the study of \cite{McCormick(2013)ApJ_774_126}.
They investigated the extraplanar PAH emission of the nearby galaxies with winds or extraplanar diffuse ionized gas, and showed the high correlations between the galactic infrared flux and the extraplanar PAH flux ($r=0.89$), and between $\Sigma_{\mathrm{SFR}}$ and the ratio of PAH emission characteristic height over galactic major axis size ($H_{\mathrm{ePAH}}/D_{25}$, $r=0.93$).
We here note that the PAH emission strength does not indicate the PAH mass, as mentioned in section \ref{intro}, since PAH emits infrared radiations absorbing \del{ultraviolet}\rev{UV} and optical radiations \citep{Draine(2003)ARA&A_41_241,Allamandola(1989)ApJS_71_733,Leger(1984)A&A_137_L5}.
It is therefore improper to directly compare their results with ours (Figure \ref{fig-scatter}).
However, $H_{\mathrm{ePAH}}$ might not be much different from $z_d^{\mathrm{thick}}$, as our (\sfrsd, \zdD) data points shows no prominent deviation from the linear relation between $\Sigma_{\mathrm{SFR}}$ and $H_{\mathrm{ePAH}}/D_{25}$ of \cite{McCormick(2013)ApJ_774_126}.
Our \sfrsd{} ($\sim 10^{-3}$ \Msyrkpcsq) is located at the lower range of their  $\Sigma_{\mathrm{SFR}}$, and thus a future study on combined targets may reveal the connection of extraplanar dust with the star-forming and star-bursting galaxies.
If we extends the targets to include the galaxies in the vicinity of intergalactic dust \citep[e.g.][]{Menard(2010)MNRAS_405_1025,Zaritsky(1994)AJ_108_1619,McGee(2010)MNRAS_405_2069,Chelouche(2007)ApJ_671_L97}, we could study the dust at whole scales, i.e.~from circumgalactic to intergalactic scales.

\section{Conclusions \label{concl}}
We modeled the extraplanar dust of six nearby galaxies, employing three dimensional radiative transfer Monte-Carlo simulation code, so as to investigate the relations with their host galaxy's property.
The targets were selected from the highly-inclined galaxies that show dust-scattered \del{ultraviolet}\rev{UV} halos \citep{Hodges-Kluck(2014)ApJ_789_131}, and the archival \galex{} FUV band images were used.
The configurations of dust and starlight sources are the same as \cite{Seon(2014)ApJ_785_L18}.
Two dust layers and one starlight source layer are assumed, and their radial and vertical distributions are all exponential.  
The FUV images of all six galaxies are well reproduced from the modeling, and we obtained several important physical parameters, such as \sfruv, $B$-band face-on optical depth, and scale-height.
\rev{
However, we found that the UV halo of three targets (IC 5249, NGC 24, and NGC 4173) can be modeled with the typical thin dust disk solely, since the contribution from the thick dust disk is too small.
These three targets were excluded from our further analyses, and we made scatter plots between several physical quantities for the rest three targets (NGC 891, NGC 3628, and UGC 11794).
Although the sample number is just three, the data points fall within a narrow range in the scatter plots between \Mex, \sfruv, \Mst, \zdD, and \sfrsd.
}
\del{\Mex{} shows high correlations with both \sfruv{} ($r=0.99$) and \Mst{} (i.e.~galaxy luminosity, $r=0.98$).
\zdD{} and \sfrsd{}, which are thought to be scale-independent quantities, show a low correlation of 0.22.
We think the high correlation between \Mex{} and \sfruv{} is meaningful, since the dynamic range of \sfruv{} is large enough ([\sfruv]$_{max}$/[\sfruv]$_{min}\sim30$); however, the dynamic range of \sfrsd{} ([\sfrsd]$_{max}$/[\sfrsd]$_{min}$) is only $\sim6$, hence we did not impart much meaning to the low correlation between \zdD{} and \sfrsd.}

\rev{
The needlessness of the thick dust disk for some galaxies reported to have UV halos by \cite{Hodges-Kluck(2014)ApJ_789_131} indicates that the sample galaxies of \cite{Hodges-Kluck(2014)ApJ_789_131} may be contaminated by the galaxies with negligible extraplanar (halo) dust.
This may thus affect the relations between the specific halo UV luminosity, the specific SFR, and the specific host-galaxy UV luminosity, they found.
Although the sample number is too small, their narrow locations over the scatter plots suggest that a common mechanism may regulate the behavior of the extraplanar dust.
We lists several possible mechanisms, discussing their ability to produce the observed diffuse-and-global UV halos.
They are the galactic radiation, the (magneto-)hydrodynamic phenomena, and the dust accretion from circumgalactic or intergalactic medium.
}

\del{
The high correlations of \Mex{} with both \sfruv{} and \Mst{} indicate the scaling relation among the three quantities.
Since the gravity scales with \Mst{}, the pushing force that elevates the extraplanar dust should also be scaled with \Mst{}, given that the extraplanar dust maintains a steady position away from the galactic plane.
The galactic radiation field scales with \Mst, and this possibility was previously studied by several authors }\delref{\citep{Ferrara(1991)ApJ_381_137,Franco(1991)ApJ_366_443,Ferrara(1990)A&A_240_259,Barsella(1989)A&A_209_349,Greenberg(1987)Nature_327_214,Pecker(1974)A&A_35_7,Pecker(1972)A&A_18_253,Chiao(1972)MNRAS_159_361}.}\del{
It was not straightforward to determine which one (\sfruv{} or galaxy luminosity) is more significant factor for increasing \Mex, because the galactic spectral shape varies among spiral galaxies }\delref{\citep[cf.][]{Dale(2009)ApJ_703_517,Rowan-Robinson(2005)AJ_129_1183,Yoshii(1988)ApJ_326_1}.}\del{
If the spectral shape difference has a negligible effect on the radiation pressure, \Mex{} would mainly depend on the galactic luminosity rather than \sfruv.
We considered other dynamical mechanisms that can produce the extraplanar dust, as proposed by }\delref{\cite{Howk(1997)AJ_114_2463}.}\del{
We think that those would be relatively minor, because their localized activities seems to have difficulties in producing the observed global-and-diffuse ultraviolet halos }\delref{\citep{Hodges-Kluck(2014)ApJ_789_131}.}\del{
Whereas the galactic radiation field has less difficulty due to its approximate plane-parallel propagation away from the galactic plane.
}

The extraplanar PAH emission study of nearby galaxies by \cite{McCormick(2013)ApJ_774_126} could not directly compared with ours, since the PAH emission does not simply mean the amount of PAH; PAH emits infrared radiation absorbing \del{ultraviolet}\rev{UV} and optical radiations \citep{Draine(2003)ARA&A_41_241,Allamandola(1989)ApJS_71_733,Leger(1984)A&A_137_L5}.
However, $H_{\mathrm{ePAH}}$ might not be much different from $z_d^{\mathrm{thick}}$, as our (\sfrsd, \zdD) data points shows no prominent deviation from the linear relation between $\Sigma_{\mathrm{SFR}}$ and $H_{\mathrm{ePAH}}/D_{25}$ of \cite{McCormick(2013)ApJ_774_126}.
Our \sfrsd{} is at the lower end of \citeauthor{McCormick(2013)ApJ_774_126}'s (\citeyear{McCormick(2013)ApJ_774_126}) $\Sigma_{\mathrm{SFR}}$ range, and hence we will be able to study the connection of extraplanar dust with the star-forming and starburst galaxies, by extending our targets to include theirs.

\del{We note that our results are not from an unbiased sample, hence unbiased sample studies are required to test and validate the results here we found.}

\acknowledgments
J.-H.S. is grateful to Yujin Yang, Dennis Zaritsky, and Ann Zabludoff for their valuable comments and discussion on the results. 

\bibliographystyle{apj}
\bibliography{refs_2015}

\clearpage

\clearpage
\begin{figure}
\center{
(a)\includegraphics[scale=0.7]{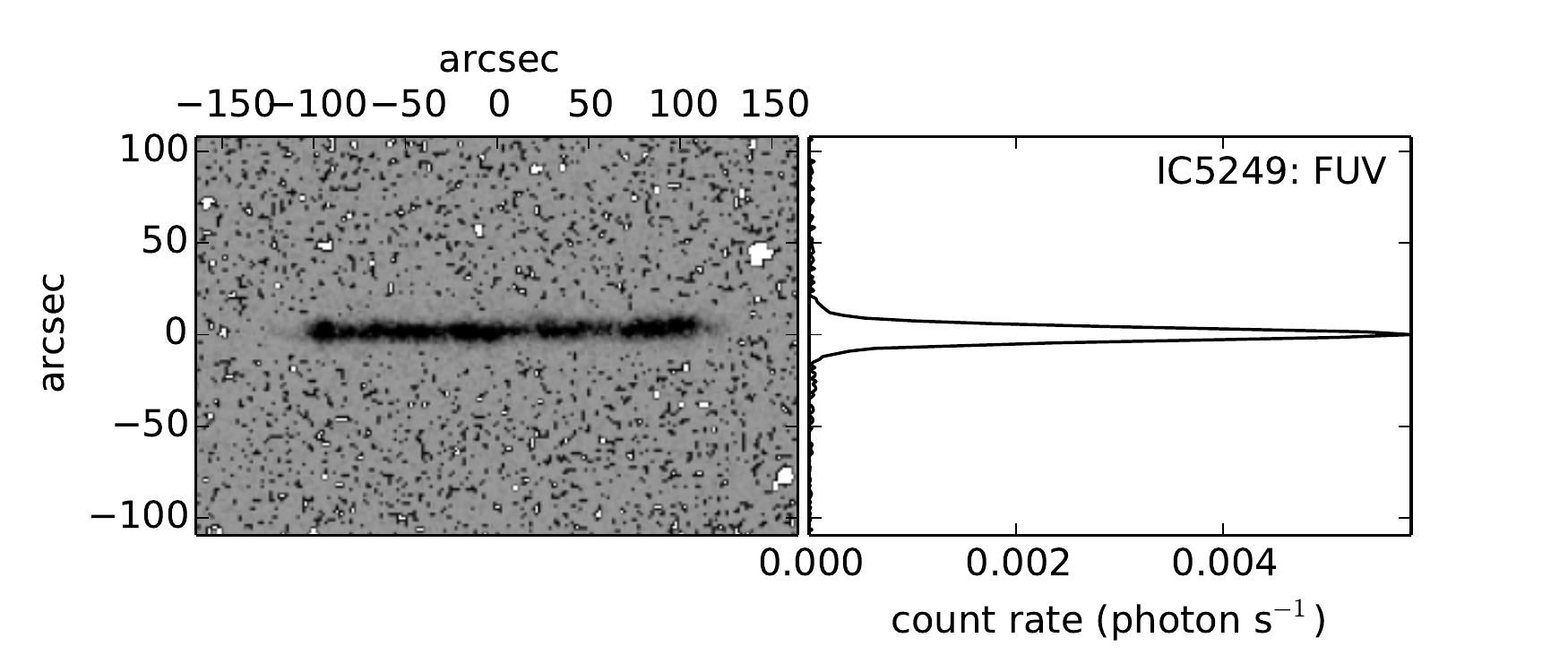}
(b)\includegraphics[scale=0.7]{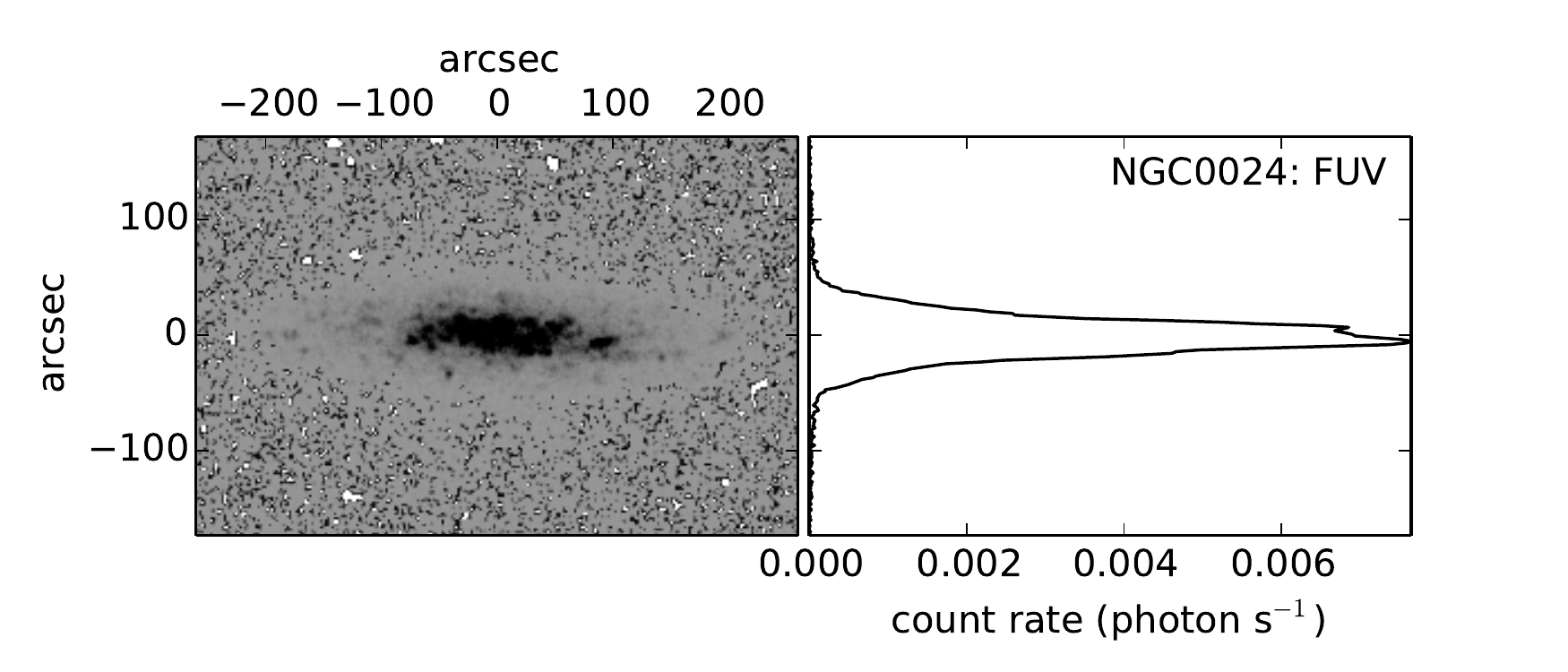}
(c)\includegraphics[scale=0.7]{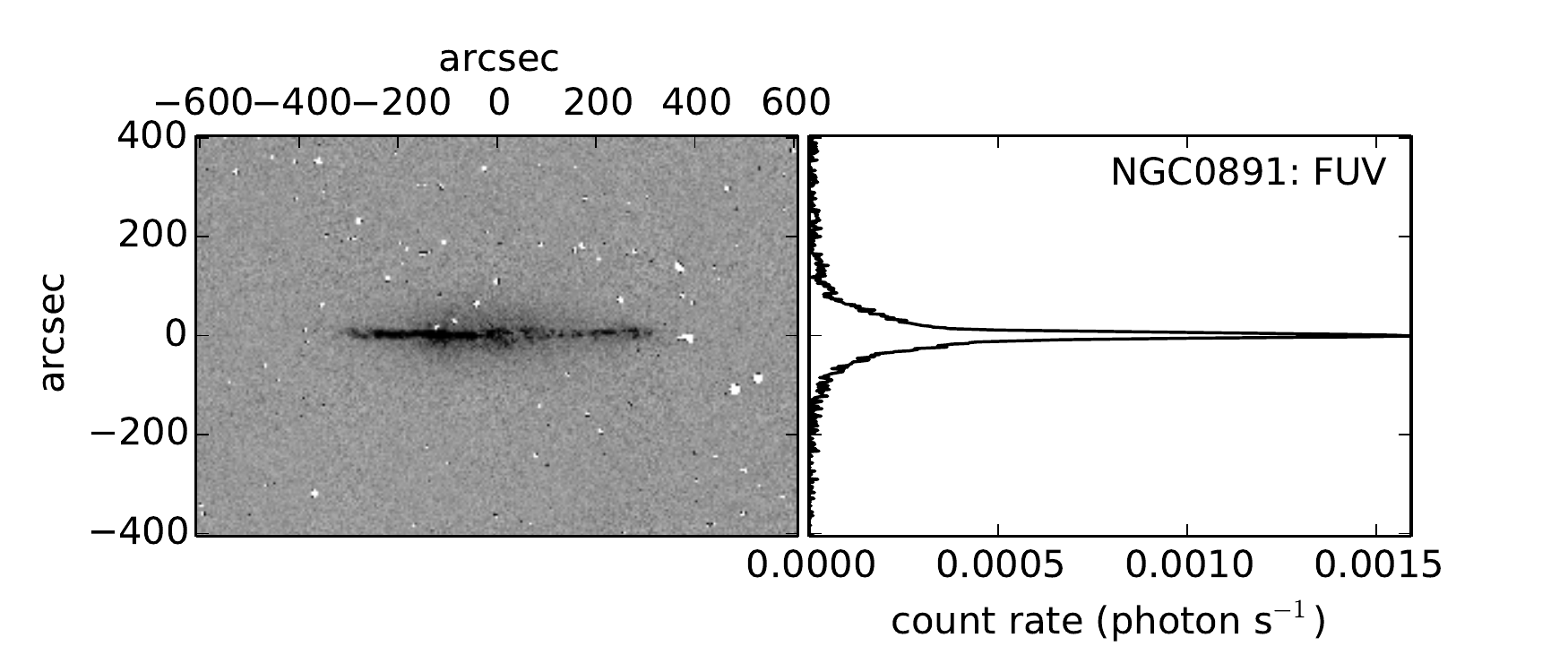}
}
\caption{\galex{} FUV images and vertical profiles of our target galaxies. \label{fig-data}}
\end{figure}

\clearpage
\begin{figure}
\figurenum{2}
\center{
(d)\includegraphics[scale=0.7]{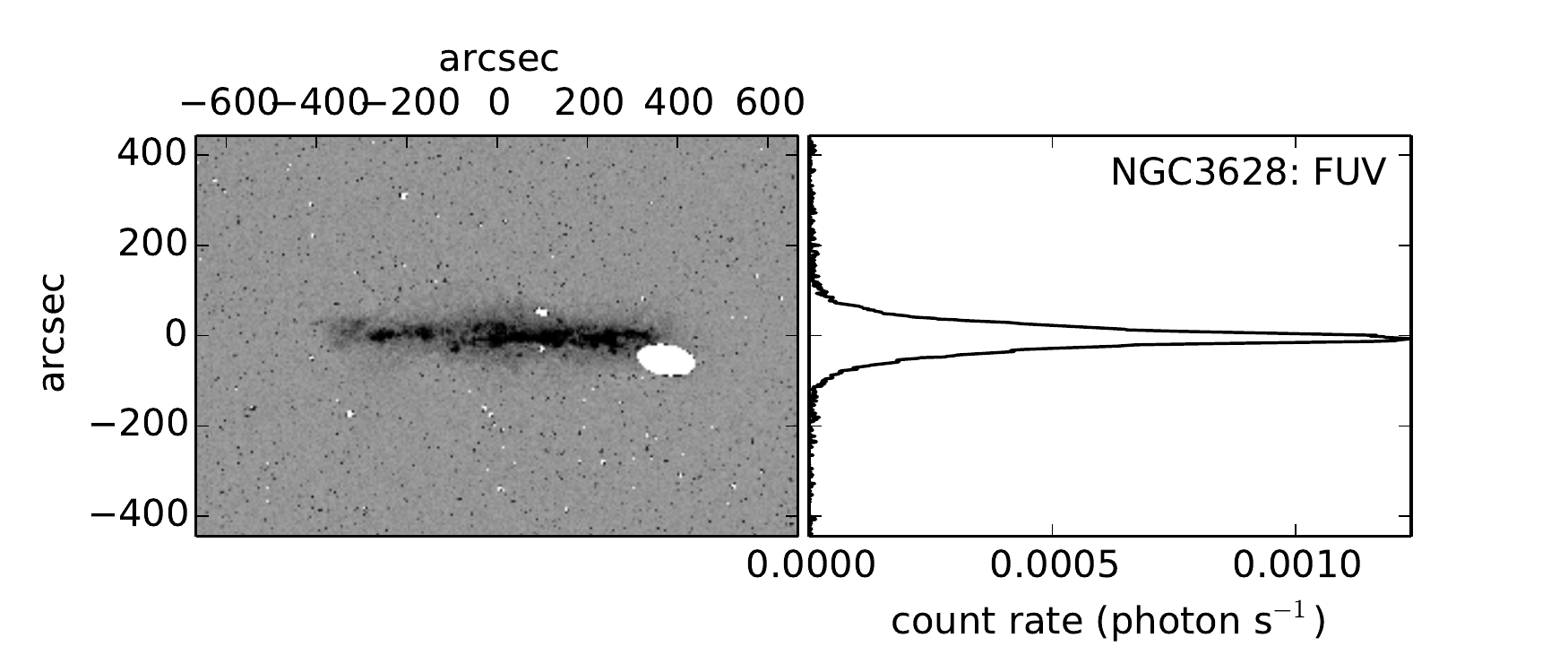}
(e)\includegraphics[scale=0.7]{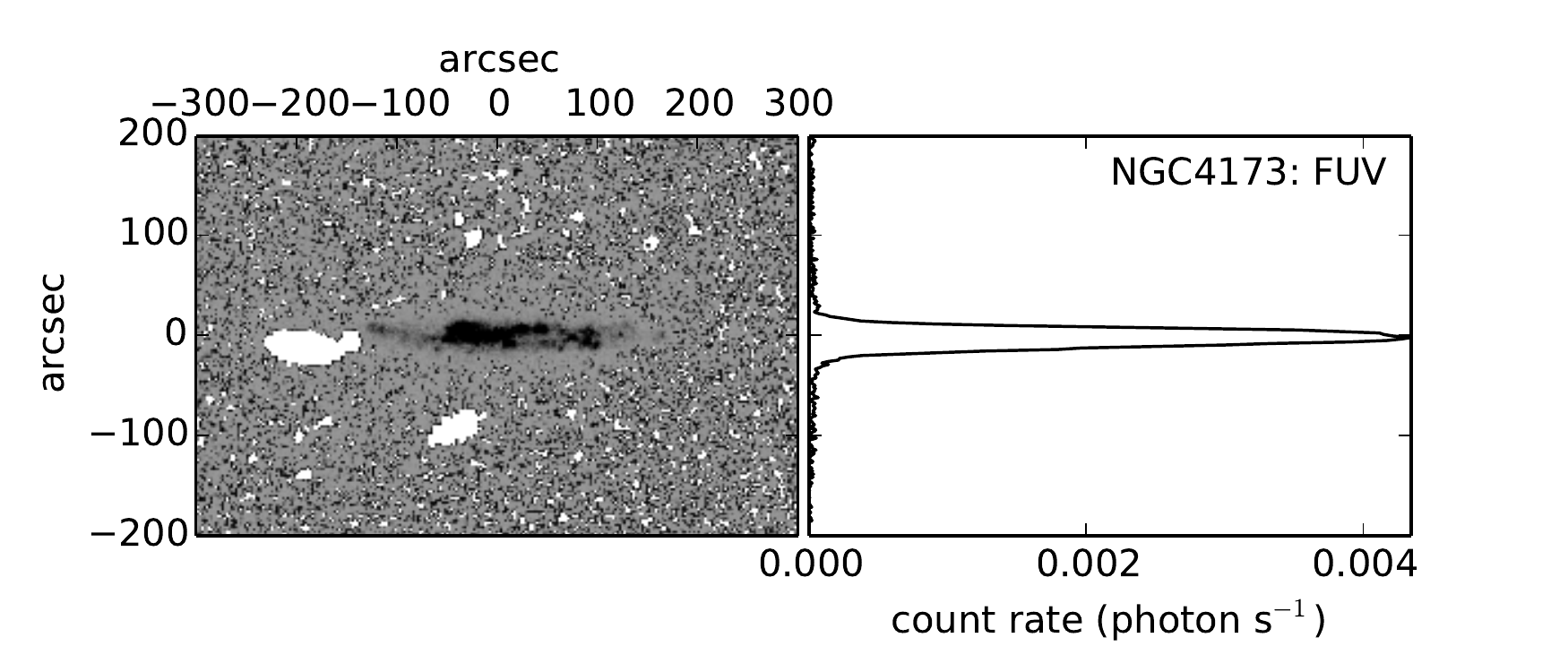}
(f)\includegraphics[scale=0.7]{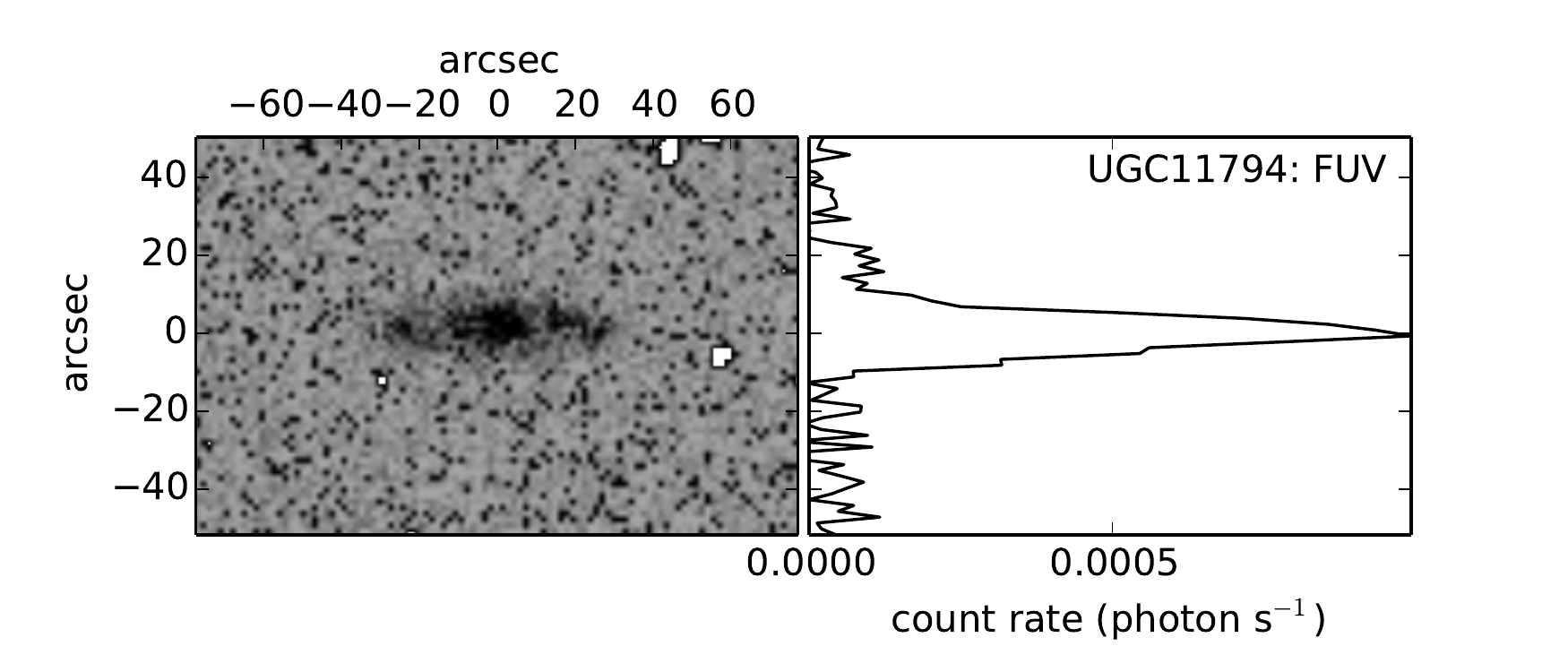}

}
\caption{Continued.}
\end{figure}


\clearpage
\begin{figure}
\figurenum{3}
\center{
(a)\includegraphics[scale=0.9,angle=90]{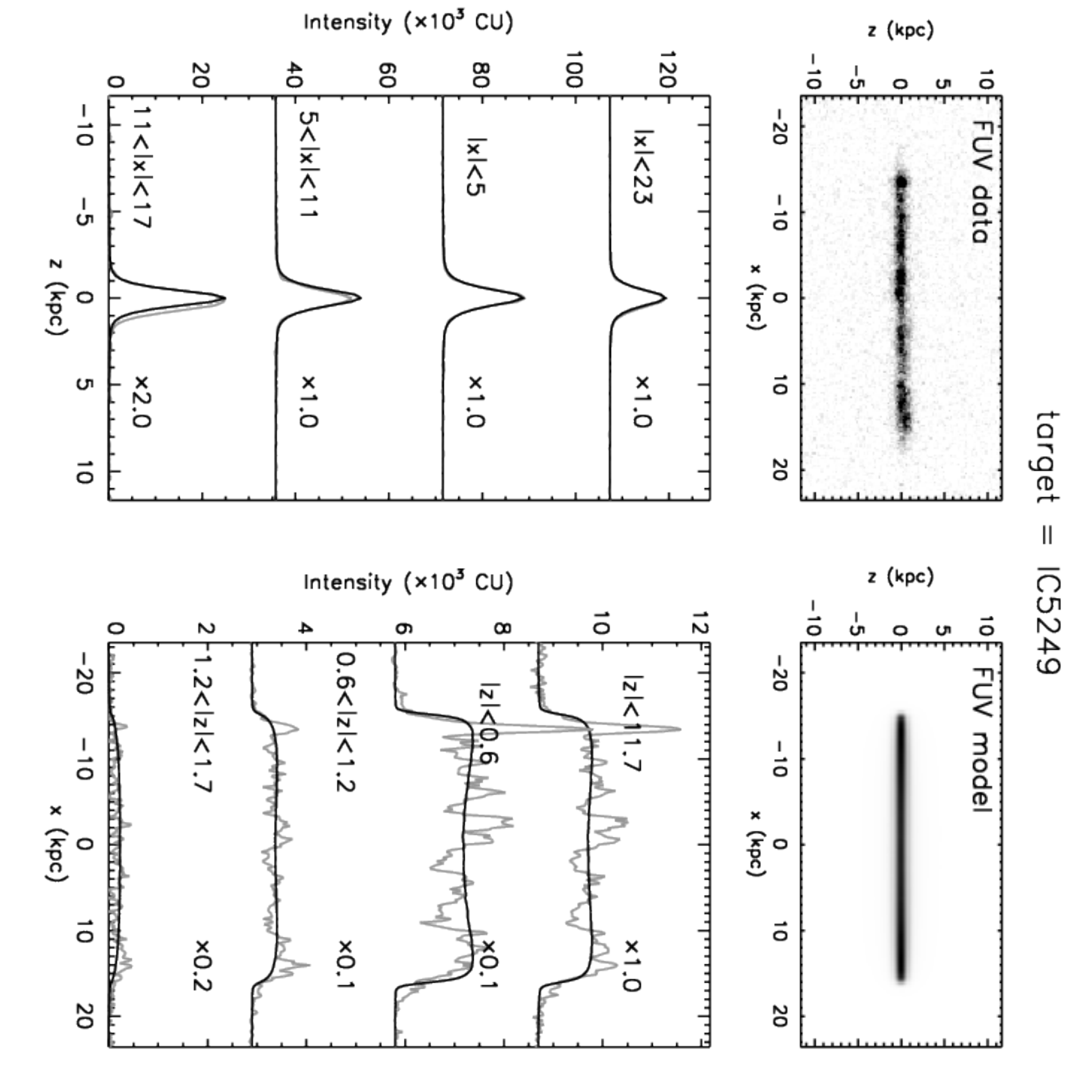}
}
\caption{Fitting results. (\textit{upper-left}) \galex{} FUV image. (\textit{upper-right}) modeled image. (\textit{lower-left}) vertical profiles (\textit{lower-right}) horizontal profiles. The profiles of the modeled image are black lines, while those of \galex{} FUV images are gray lines. (a)-(f) best-fit results for individual targets. \label{fig-fit}}
\end{figure}

\clearpage
\begin{figure}
\figurenum{3}
\center{
(b)\includegraphics[scale=0.9,angle=90]{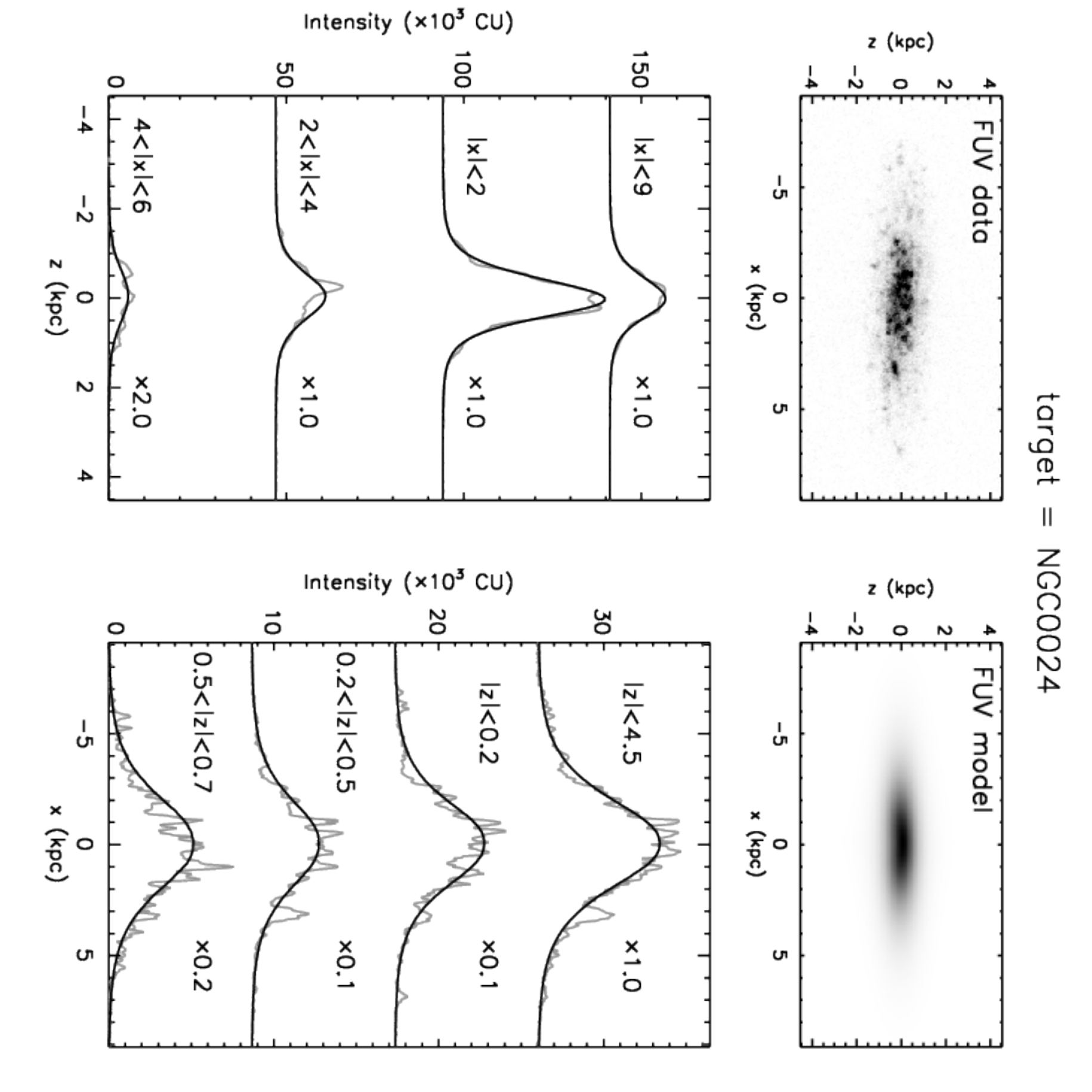}
}
\caption{Continued.}
\end{figure}

\clearpage
\begin{figure}
\figurenum{3}
\center{
(c)\includegraphics[scale=0.9,angle=90]{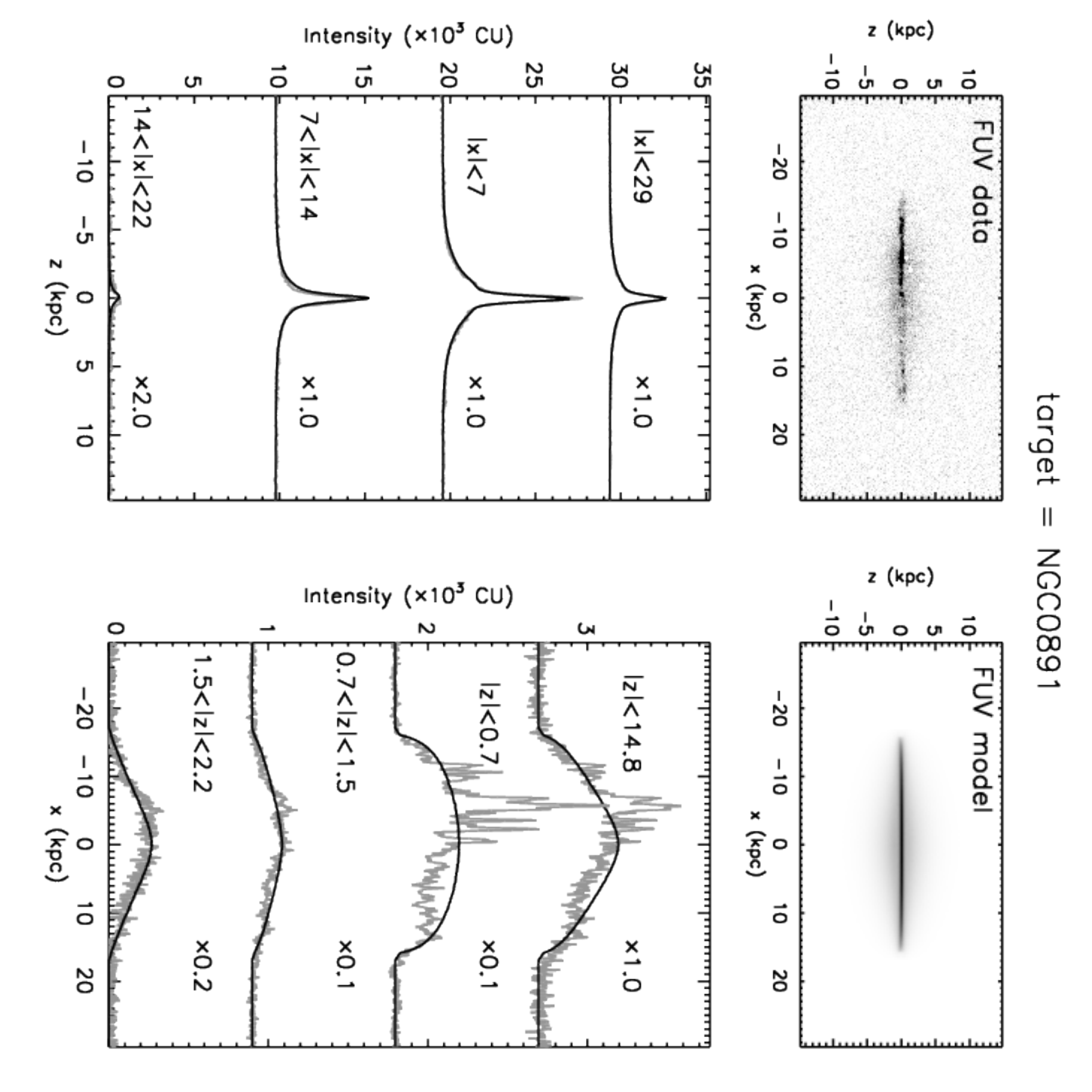}
}
\caption{Continued.}
\end{figure}

\clearpage
\begin{figure}
\figurenum{3}
\center{
(d)\includegraphics[scale=0.9,angle=90]{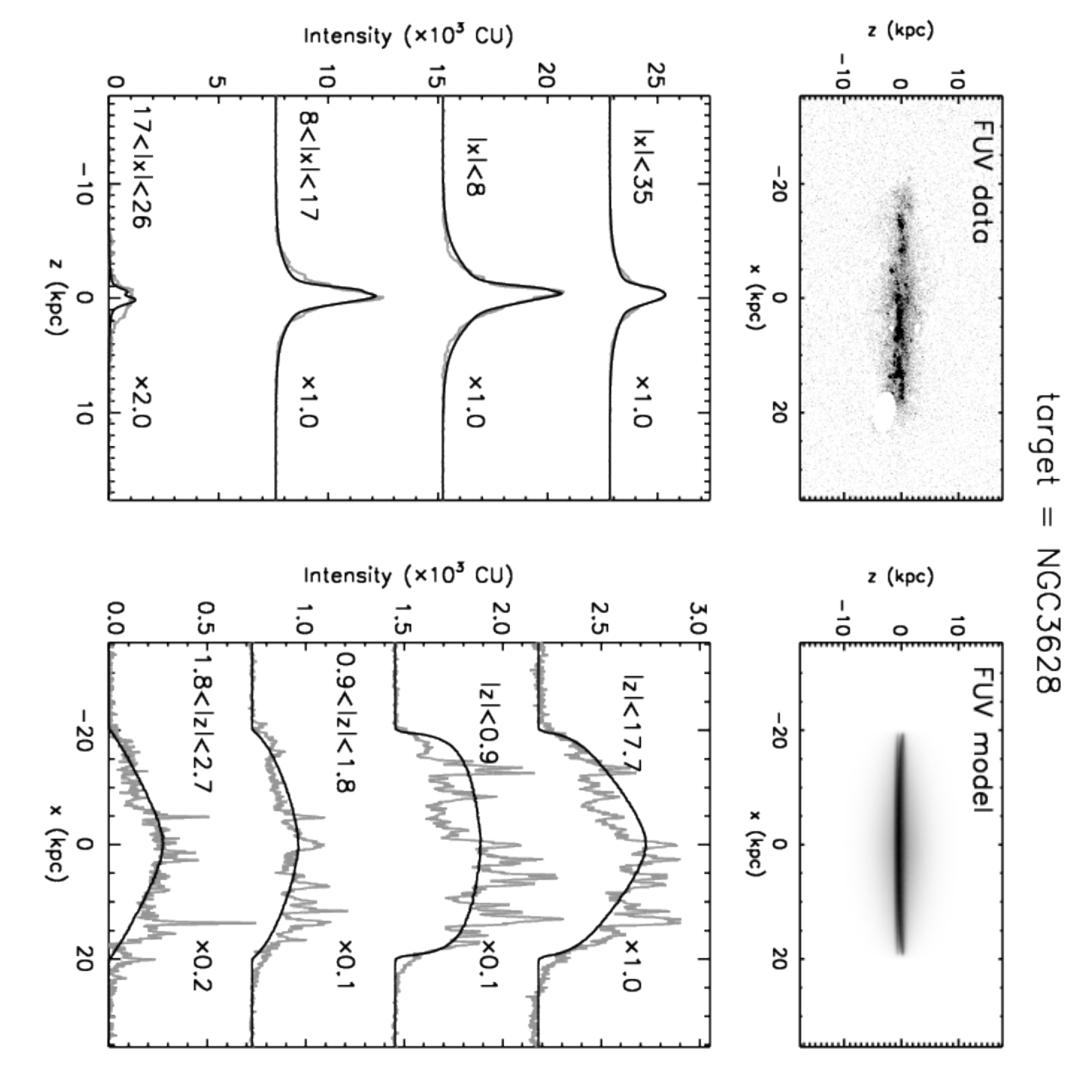}
}
\caption{Continued.}
\end{figure}

\clearpage
\begin{figure}
\figurenum{3}
\center{
(e)\includegraphics[scale=0.9,angle=90]{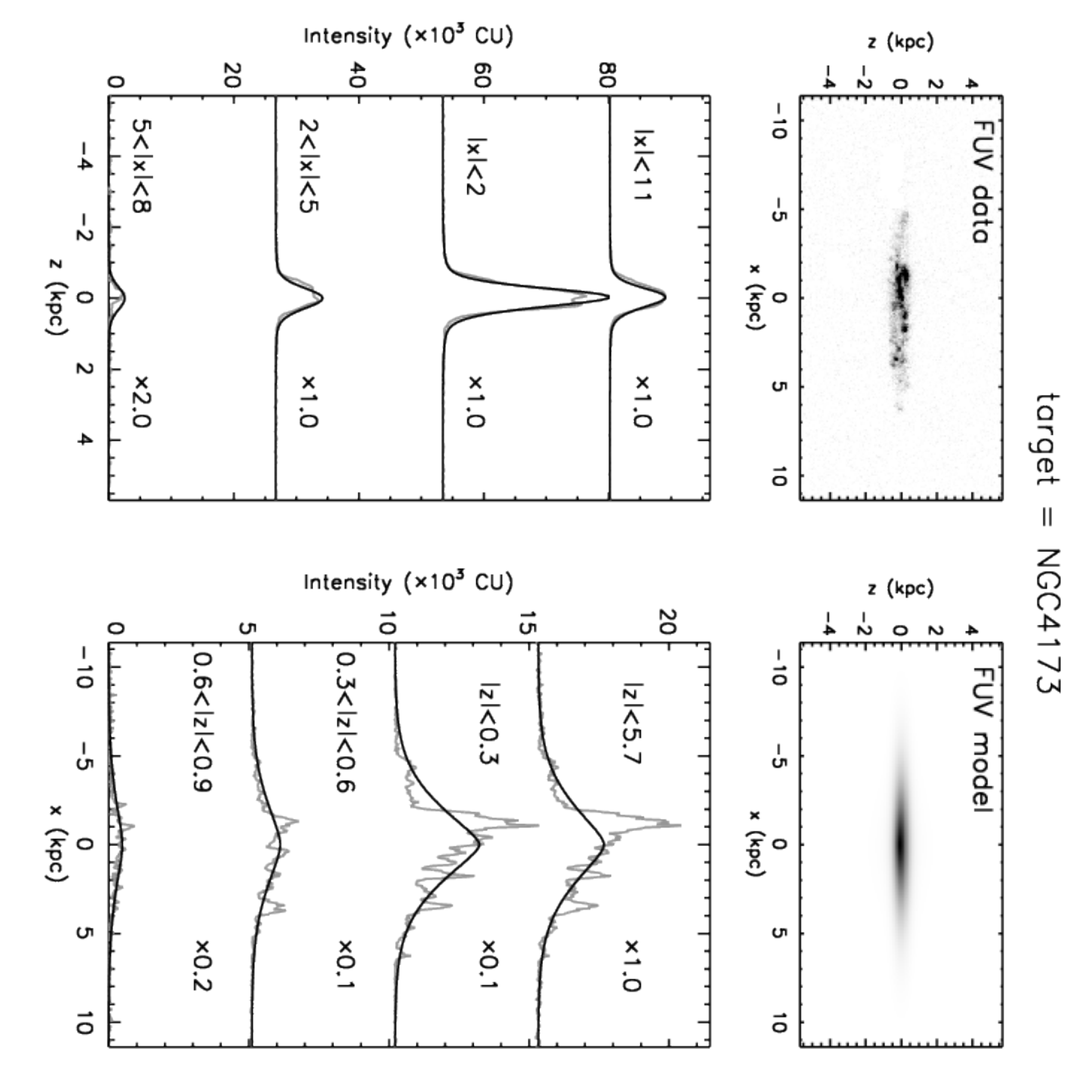}
}
\caption{Continued.}
\end{figure}

\clearpage
\begin{figure}
\center{
(f)\includegraphics[scale=0.9,angle=90]{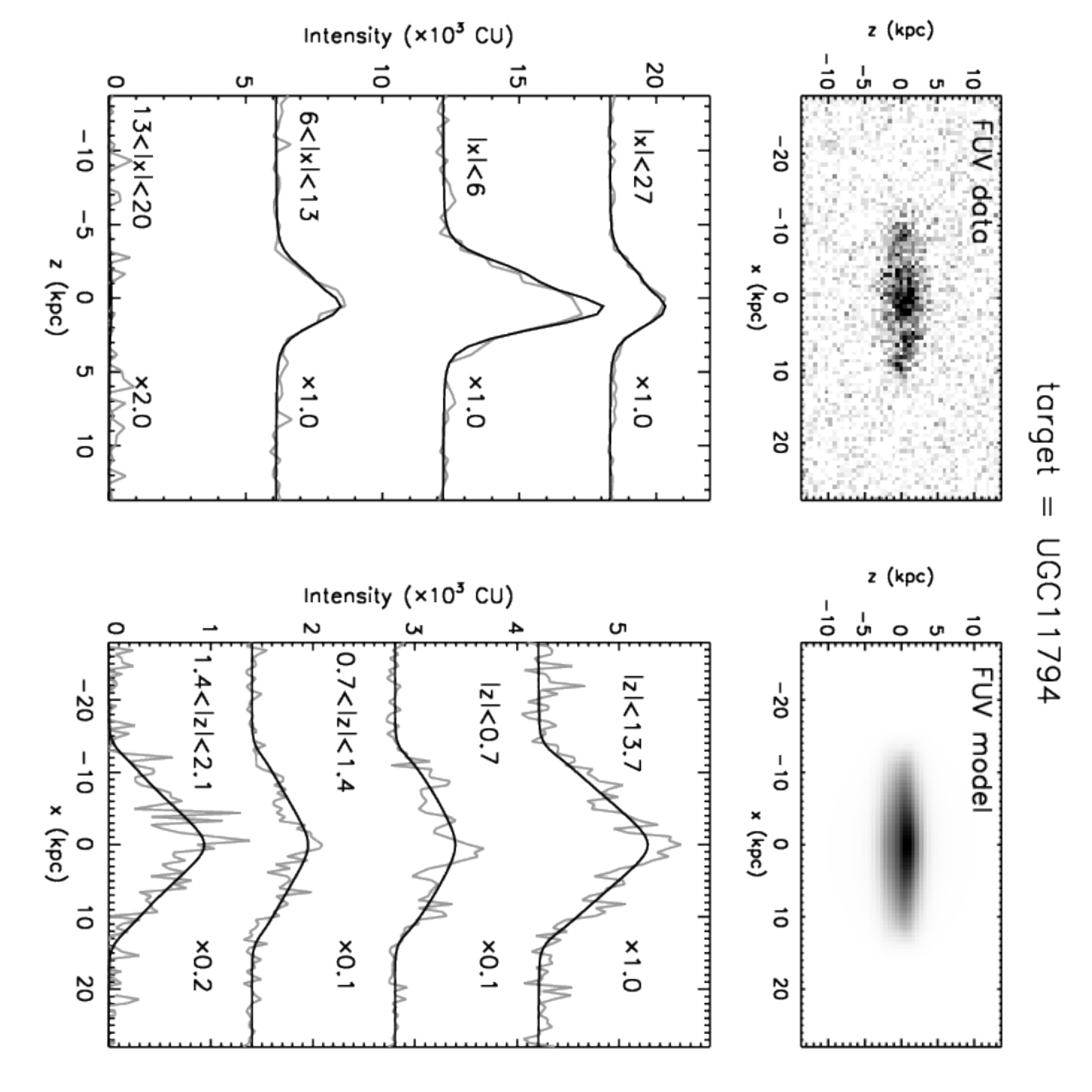}
}
\caption{Continued.}
\end{figure}

\clearpage

\clearpage

\clearpage
\begin{figure}
\center{
\includegraphics[scale=0.7,angle=0]{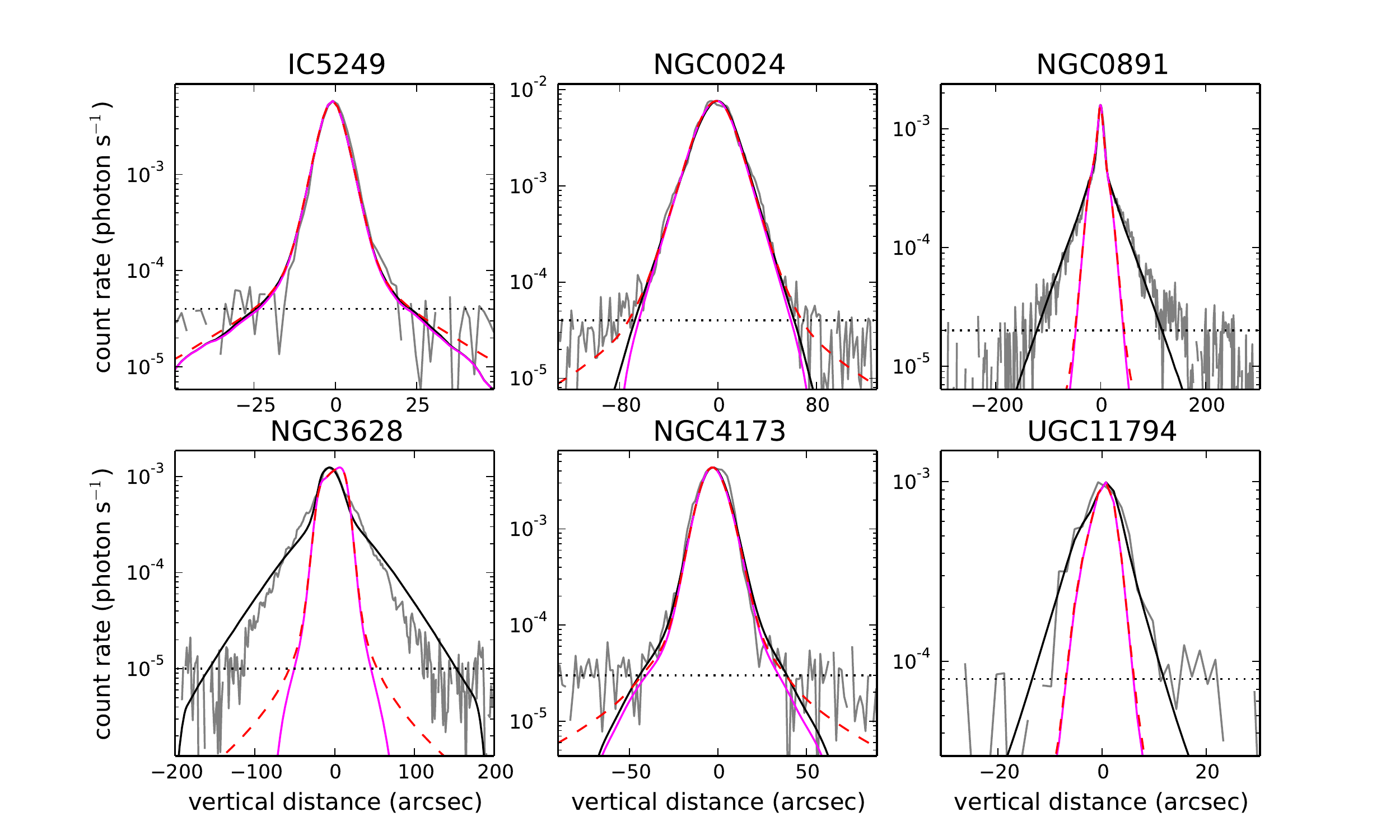}
}
\caption{The effects of the extended PSF on the model fitting. \rev{All lines are the horizontally averaged vertical profiles. The \textit{gray-solid line} is from the \galex{} FUV data. The \textit{black-solid line} is from the best-fit model convoluted with the \textit{standard} PSF. The \textit{magenta-solid and red-dashed lines} are from the best-fit model convoluted with the \textit{standard} and \textit{extended} PSF, respectively, excluding the thick dust disk component (see eq.~(\ref{eq-kappa})). The \textit{dotted line} indicates the upper bound of background noise.} \del{The \textit{black line} is the horizontally-averaged vertical profile of the model difference, i.e.~(best-fit model convoluted with the \textit{extended} PSF) $-$ (best-fit model convoluted with the \textit{standard} PSF). The \textit{gray line} is for the difference of (data) $-$ (best-fit model convoluted with the \textit{standard} PSF). Note the much smaller y-axis scale, compared to the one in Figure \ref{fig-data}.} \label{fig-diff}}
\end{figure}

\clearpage

\clearpage
\begin{deluxetable}{cccrrrrr}
\tablewidth{0pt}
\tablecaption{Target information \label{tbl-tg}}
\tablehead{
\colhead{Target} &\colhead{Type} &\colhead{$i$} &\colhead{$d$} &\colhead{$D_{25}$} &\colhead{$D_{25}$} &\colhead{$M_*$} &\colhead{SFR$_{\mathrm{IR}}$}\\
&	&\colhead{($\arcdeg$)} 	&\colhead{(Mpc)}	&\colhead{($\arcmin$)} &\colhead{(kpc)}	&\colhead{($10^{10}\,M_\odot$)} &\colhead{(\Msyr)}\\
&\colhead{(1)}	&\colhead{(2)} 	&\colhead{(3)}	&\colhead{(4)}	&\colhead{(5)} &\colhead{(6)} &\colhead{(7)}
}
\startdata
IC 5249       &SBcd  &       90 &    29.70 &     3.63 &    31.36 &     0.15 &      \nodata\\
NGC 0024      &Sc    &       70 &     7.24 &     5.75 &    12.12 &     0.15 &     0.11\\
NGC 0891      &Sb    &       90 &    10.06 &    13.49 &    39.48 &     5.00 &      2.4\\
NGC 3628      &Sb    &       79 &    10.95 &    14.79 &    47.11 &     7.00 &      4.8\\
NGC 4173      &SBcd  &       90 &     7.83 &     5.01 &    11.41 &      \nodata &      \nodata\\
UGC 11794     &Sab   &       78 &    75.40 &     1.20 &    26.36 &      \nodata &$<$      5.2\\
\enddata
\tablecomments{
Columns: 
(1) morphological type, 
(2) disk inclination angle, 
(3) distance to the target, 
(4)-(5) major axis size, 
(6) stellar mass, and
(7) the star formation rate estimated from the infrared luminosity.
Columns (1)-(2) are taken from the HyperLeda galaxy database, \url{http://leda.univ-lyon1.fr}. 
Columns (3)-(4) are taken from the NED database, \url{http://ned.ipac.caltech.edu/}.
Column (5) is calculated with Columns (3) and (4).
Columns (6)-(7) are taken from \cite{Hodges-Kluck(2014)ApJ_789_131}. Column (6) is computed from the absolute $K$-band magnitude of 2MASS catalog \citep{Skrutskie(2006)AJ_131_1163} and the color corrections from \cite{Bell(2001)ApJ_550_212}. Column (7) is estimated from the relation between SFR and IR luminosity \citep{Kennicutt(1998)ApJ_498_541}. The IR luminosity is measured using the \cite{Rice(1988)ApJS_68_91} relation and the \anchor{http://irsa.ipac.caltech.edu/Missions/iras.html}{\textit{IRAS}} catalog.
}
\end{deluxetable}

\clearpage
\begin{deluxetable}{cccccc}
\tablewidth{0pt}
\tablecaption{\galex{} information \label{tbl-galex}}
\tablehead{
\colhead{Band} &  \colhead{Bandwidth} & \colhead{Effective wavelength} & \colhead{Field of view} & \colhead{Image resolution} & \colhead{Image pixel scale}\\
&\colhead{(\AA)} 	&\colhead{(\AA)}	&\colhead{($\arcdeg$)}	&\colhead{($\arcsec$)}	&\colhead{($\arcsec$)}
}
\startdata
FUV	&1344--1786	&1538.6 &1.27	&4.2	&1.5\\
NUV	&1771--2831 &2315.7 &1.25	&5.3	&1.5\\
\enddata
\tablecomments{All the information is from \cite{Morrissey(2007)ApJS_173_682}.}
\end{deluxetable}

\clearpage
\begin{deluxetable}{ccc}
\tablewidth{0pt}
\tablecaption{\galex{} Exposure Time \label{tbl-exp}}
\tablehead{
\colhead{Target} &  \colhead{FUV} & \colhead{NUV}\\ 
&\colhead{(s)} 	&\colhead{(s)}
}
\startdata
IC 5249       &     1705 &     4467\\
NGC 0024      &     1577 &     1577\\
NGC 0891      &     6047 &     6283\\
NGC 4173      &     1648 &     1648\\
NGC 7090      &     3007 &     3088\\
\enddata
\end{deluxetable}

\clearpage
\begin{deluxetable}{crrrrrrrrrrrr}
\tablewidth{0pt}
\tabletypesize{\scriptsize}
\tablecaption{Fitting Parameters of the Best Fit for the \galex{} FUV Images \label{tbl-fit}}
\tablehead{
\colhead{Target} &\colhead{\tablenotemark{a}$z_d^{\mathrm{thin}}$} &\colhead{\tablenotemark{a}$z_d^{\mathrm{thick}}$} &\colhead{$\tau_B^{\mathrm{thin}}$} &\colhead{$\tau_B^{\mathrm{thick}}$} &\colhead{$z_s$} &\colhead{$h_d$} &\colhead{$h_s$} &\colhead{$R_d$} &\colhead{\tablenotemark{b}$Z_d$} &\colhead{$R_s$} &\colhead{$i$} &\colhead{SFR$_{\mathrm{UV}}$} \\
&\colhead{(kpc)} &\colhead{(kpc)} & & &\colhead{(kpc)} &\colhead{(kpc)} &\colhead{(kpc)} &\colhead{(kpc)} &\colhead{(kpc)} &\colhead{(kpc)} &\colhead{($\arcdeg$)} &\colhead{(\Msyr)} \\
&\colhead{(1)}	&\colhead{(2)} 	&\colhead{(3)}	&\colhead{(4)}	&\colhead{(5)}  &\colhead{(6)}	&\colhead{(7)} 	&\colhead{(8)}	&\colhead{(9)}	&\colhead{(10)} &\colhead{(11)} &\colhead{(12)}
}
\startdata
IC5249 &0.375 &2.600 &0.562 &0.013 &0.208 &9.3 &122.5 &15.7 &5.0 &15.7 &90.2 &0.57\\
 &&&(0.333) &(0.218) &(0.025) &(1.0) &(52.5) &(0.7) &&(0.5) &(0.3) &(0.19)\\
NGC0024 &0.175 &3.100 &0.443 &1.002 &0.146 &0.8 &1.3 &12.5 &5.0 &12.5 &78.1 &0.079\\
 &&&(0.125) &(0.778) &(0.037) &(0.1) &(0.1) &(2.8) &&(4.1) &(0.6) &(0.0036)\\
NGC0891 &0.325 &2.100 &0.985 &0.436 &0.102 &12.0 &5.8 &16.9 &12.0 &15.9 &89.5 &1.4\\
 &&&(0.412) &(0.285) &(0.014) &(2.0) &(0.9) &(3.2) &&(1.6) &(0.4) &(0.34)\\
NGC3628 &0.225 &2.100 &1.230 &0.640 &0.164 &16.4 &7.8 &20.0 &10.0 &19.5 &91.6 &2.4\\
 &&&(1.115) &(0.267) &(0.127) &(22.1) &(13.6) &(6.9) &&(1.5) &(2.0) &(1.3)\\
NGC4173 &0.125 &0.600 &0.371 &0.093 &0.084 &2.8 &1.2 &19.7 &5.0 &19.2 &85.0 &0.082\\
 &&&(0.365) &(0.469) &(0.373) &(2.9) &(1.9) &(13.1) &&(10.8) &(6.6) &(0.020)\\
UGC11794 &0.175 &1.600 &0.747 &0.450 &0.201 &15.3 &4.2 &18.3 &10.0 &13.8 &91.1 &1.9\\
 &&&(0.686) &(0.227) &(0.174) &(6.2) &(2.9) &(2.5) &&(2.9) &(8.0) &(1.7)\\

\enddata
\tablecomments{\rev{The numbers surrounded by parenthesis in the table body are the corresponding uncertainties derived from the area half $z$-step away from the best-fit ($z_d^{\mathrm{thin}}$,$z_d^{\mathrm{thick}}$).}
Columns: 
(1) the scale height of the thin dust disk, 
(2) the scale height of the thick dust disk, 
(3) the $B$-band optical depth along the symmetric axis of the thin dust disk, 
(4) the $B$-band optical depth along the symmetric axis of the thick dust disk, 
(5) the scale height of the stellar disk, 
(6) the scale length of the thin and thick dust disks, 
(7) the scale length of the stellar disk, 
(8) the radial extent of the thin and thick dust disks,
(9) the vertical extent of the thin and thick dust disks,
(10) the radial extent of the stellar disk, 
(11) the disk inclination angle measured from the top of the image downward (cf.~Figure \ref{fig-data}), and
(12) the star formation rate estimated from the intrinsic ultraviolet luminosity.}
\tablenotetext{a}{We covered several ($z_d^{\mathrm{thin}}$, $z_d^{\mathrm{thick}}$) grids. See text for detail.}
\tablenotetext{b}{This parameter is fixed during the fitting.}
\end{deluxetable}

\clearpage
\begin{deluxetable}{ccc}
\tablewidth{0pt}
\tablecaption{Adopted $z_x$ and Estimated \Mex{} \label{tbl-zxMext}}
\tablehead{
\colhead{Target} &  \colhead{$z_x$} & \colhead{\Mex}\\ 
&\colhead{(kpc)} 	&\colhead{(\Ms)}
}
\startdata
NGC0891       &     1.97 &     8.91e+06\\
NGC3628       &     2.36 &     1.64e+07\\
UGC11794      &     1.32 &     1.35e+07\\
\enddata
\end{deluxetable}

\end{document}